\documentclass[aps,pre,preprint]{revtex4-1} 
\usepackage{amsfonts}
\usepackage{amsmath}
\usepackage{amssymb}
\usepackage{graphicx}
\usepackage{mathrsfs} 
\usepackage{color} 

\providecommand\bnabla{\boldsymbol{\nabla}}

\providecommand\bcdot{\boldsymbol{\cdot}}

\providecommand\bx{\mathbf{x}}

\providecommand\bn{\mathbf{\hat{n}}}

\newcommand\dd{d} 

\newcommand{\pd}[2]{\frac{\partial #1}{\partial #2}}

\newcommand{\bb}[1]{\boldsymbol{#1}}
\newcommand{\ub}[1]{^{({#1})}}

\begin{document} 
\title{Asymptotic approximations for the plasmon\\ resonances of nearly touching spheres}
\author{Ory Schnitzer}
\affiliation{Department of Mathematics, Imperial College London, South Kensington Campus, London SW7 2AZ, United Kingdom}

\begin{abstract}
Excitation of surface-plasmon resonances of closely spaced nanometallic structures is a key technique used in nanoplasmonics to control light on subwavelength scales and generate highly confined electric-field hotspots. In this paper we develop asymptotic approximations in the near-contact limit for the entire set of surface-plasmon modes associated with the prototypical sphere dimer geometry. Starting from the quasi-static plasmonic eigenvalue problem, we employ the method of matched asymptotic expansions between a gap region, where the boundaries are approximately paraboloidal, pole regions within the spheres and close to the gap, and a particle-scale region where the spheres appear to touch at leading order. For those modes that are strongly localised to the gap, relating the gap and pole regions gives a set of effective eigenvalue problems formulated over a half space representing one of the poles. We solve these problems using integral transforms, finding asymptotic approximations, singular in the dimensionless gap width, for the eigenvalues and eigenfunctions. In the special case of modes that are both axisymmetric and odd about the plane bisecting the gap, where matching with the outer region introduces a logarithmic dependence upon the dimensionless gap width, our analysis follows [O. Schnitzer, \textit{Physical Review B}, \textbf{92} 235428 2015]. We also analyse the so-called anomalous family of even modes, characterised by field distributions excluded from the gap. We demonstrate excellent agreement between our asymptotic formulae and exact calculations.
\end{abstract}
\maketitle

\section{Introduction}
\subsection{The plasmonic eigenvalue problem}
The unique optical properties of metal nanoparticles and nanostructures at visible frequencies enable guiding, localisation and enhancement of light on nanometric scales, below the so-called diffraction limit, with applications to bio-sensing, photovoltaics, medical treatment, optical circuitry and metamaterials \cite{Maier:07,Engheta:07,Anker:08,Sperling:08,Schuller:10,Atwater:10}. These applications rely on the resonant excitation of localised surface plasmons, namely standing-wave oscillations of the electron-charge distribution at the metal-dielectric boundary and its induced electric field. Resonances are excited by external forcing, e.g., far-field illumination or near-field sources, at a frequency where the metal's complex permittivity is close to a resonant value. Theoretically, when the metal's permittivity is exactly equal to a resonant value, the structure can sustain a localised-plasmon oscillation in the absence of any external forcing. In that case, the forced response could potentially diverge. In reality, however, the resonant values can never be exactly attained, and accordingly the resonance is always damped. In particular, in the quasi-static limit (structures small compared with the wavelength) resonant permittivities are negative real. In comparison, while the permittivity of a metal has a negative real part at frequencies below the plasma frequency, it also has an imaginary part owing to ohmic losses. 

The quasi-static surface-plasmon modes of a nanometallic structure are governed by the plasmonic eigenvalue problem, which has recently received a lot of renewed attention in both physics and mathematics \cite{Bergman:79,Ouyang:89,Fredkin:03,Mayergoyz:05,Grieser:09,Mayergoyz:Book,Sandu:13,Grieser:14,Klimov:14,Ando:16,Davis:17,Voicu:17,Ammari:17,Bergman:18}. The problem naturally arises by considering the near-field limit of the macroscopic Maxwell equations governing the time-harmonic electric field in the vicinity of the structure, in the absence of external forcing. In this limit, the electric field is irrotational and can therefore be described by a potential, $\varphi(\bx;\omega)$, where $\bx$ is the position vector and $\omega$ is the angular frequency. This potential satisfies
\begin{equation}\label{ev1}
\bnabla\bcdot(\varepsilon(\bx;\omega)\,\bnabla\varphi) = 0,
\end{equation}
where $\varepsilon(\bx;\omega)\equiv \epsilon(\omega)$ within the bounded, possible multiply connected, domain of the structure, while $\varepsilon(\bx; \omega)\equiv 1$ in the surrounding background; note that $\epsilon(\omega)$ is the permittivity of the metal nanostructure relative to that of the background. Requiring the field $-\bnabla\varphi$ to attenuate at large distances ensures matching with outward radiating solutions of Maxwell's equations (considered on the much larger scale of the wavelength). Clearly, a field vanishing for all $\bx$ is always a solution of the above quasi-static problem. For certain special values of $\epsilon$, however, there exist, in addition, one or more non-trivial solutions. The plasmonic eigenvalue problem is thus to find all such pairs of $\epsilon$ eigenvalues and corresponding eigen-fields.

The plasmonic eigenvalue problem does not involve the frequency or any material parameters. It is also scale invariant, i.e., the uniform scaling $\bx\rightarrow l\bx$ leaves the eigenvalues $\epsilon$ unchanged, while the eigenfunctions transform as $\varphi(\bx)\rightarrow\varphi(\bx/l)$. Thus the eigenvalue problem is of a purely geometric nature, depending only on the \emph{shape} of the metallic nanostructure. Inclusions with smooth boundaries possess an infinite discrete set of negative-real $\epsilon$ eigenvalues accumulating at $\epsilon=-1$ (the latter is the condition for quasi-static surface waves at a flat metal-dielectric interface). 

Let us assume that the plasmonic eigenvalue problem has been solved for a given shape. Then the optical response of an actual metallic structure having that shape, subject to an arbitrary distribution of external sources, is readily obtained as an explicit combination of the eigenmodes. This spectral representation, which is based on orthogonality and completeness properties of the eigenfunctions, is especially efficient close to resonance where typically one or a few excited surface-plasmon modes dominate the sum \cite{Bergman:03,Klimov:07dipole,Klimov:14,Davis:17,Bergman:18}. We emphasise that, in contrast to the permittivity eigenvalues, the dielectric function of the actual metallic structure is frequency dependent and complex valued. 

\subsection{Motivation and goals}
Surface-plasmon resonance is generally manifested by amplification of the electric near-field, absorption within the metal structure and radiation away from it. It has been widely demonstrated that these features can be greatly enhanced using nanostructure geometries characterised by disparate length scales, e.g., closely spaced particles, particles nearly touching a substrate, as well as elongated particles. In particular, closely spaced structures have been extensively used to generate giant field enhancements in highly confined hotspots, at resonant frequencies controlled by the clearance \cite{Nordlander:04,Gunnarsson:05,Romero:06,Muskens:07,Hill:10,Schuller:10,Lei:12,Chen:13}. This phenomenon is inherent to nanoplasmonic applications based on optical nonlinearities, targeted heating and stimulated emission \cite{Huang:06,Sperling:08,Giannini:11,Kauranen:12}. 

In light of their practical importance, closely spaced plasmonic structures have been the subject of numerous theoretical investigations. Many of these studies are based on generic numerical methods, such as finite-element simulations, the discrete-dipole method and multipole expansions, that are computationally expensive and often struggle when the geometric disparity is very strong. Otherwise, a wide range of analytical techniques have been used, including separation of variables and transformation optics. In three dimensions, however, exact analytical methods are limited to idealised geometries and yield numerical schemes rather than closed-form formulae. While the geometric disparity is not \textit{a priori} encoded in the methods, they have been used to generate robust numerical schemes and in some cases analytical approximations in the near-contact limit \cite{Klimov:07,Klimov:07cluster,Lebedev:10,Lebedev:13,Pendry:13,Klimov:14,Yu:18}. The latter indirect approach of obtaining near-contact approximations is technical and difficult to generalise to non-ideal geometries or more sophisticated physics.

In this paper we adopt an alternative asymptotic approach where the near-contact limit is considered from the outset using matched asymptotic expansions \cite{Hinch:91}. 
Previously, we applied matched asymptotics to the problem of closely spaced metallic spheres illuminated by a plane wave, the electric field being polarised along the line of centres \cite{Schnitzer:15plas}. In particular, asymptotic approximations were obtained for the family of modes excitable in that scenario: axisymmetric modes with polarisation-charge and potential distributions odd about the plane bisecting the gap. Based on these eigensolutions, asymptotic formulae for the resonant field enhancements in the gap were derived and compared with numerical solutions. Our approach was later also applied to a generalised plasmonic description based on the hydrodynamic Drude model \cite{Schnitzer:16,Schnitzer:16b}, considering closely spaced cylinders and spheres among other shapes. These works demonstrate, in the context of nanoplasmonics, some of the typical advantages of using matched asymptotics to study singular limits. In particular, the method furnishes asymptotic formulae, e.g., for eigenvalues and field enhancements, in conjunction with a physically descriptive picture that illuminates scalings and dominant physical mechanisms. Furthermore, results are subject to considerable generalisation as the analysis does not rely on the existence of a detailed exact solution.

The axisymmetric odd modes considered in \cite{Schnitzer:15plas} constitute just one family of modes among the many comprising the remarkably rich spectrum of the sphere-dimer geometry. The modes are usually catalogued into several families based on symmetries, how well they couple with different near- and far-field external sources, as well as their asymptotic behaviour in the near-contact limit \cite{Klimov:07,Pendry:13,Klimov:14}. In this paper we will use matched asymptotic expansions to develop asymptotic approximations in the near-contact limit for the entire set of plasmonic eigenvalues and eigenfunctions of a pair of identical spheres. 

In section \S\ref{sec:formulation} we formulate the plasmonic eigenvalue problem for a sphere dimer. In \S\ref{sec:limit} we prepare for the use of matched asymptotic expansions in the near-contact limit by describing the three asymptotic regions that take part in the analysis. The analysis is carried out in \S\ref{sec:odd}--\S\ref{sec:anom}, each section addressing an asymptotically distinct family of modes; for completion we review in \S\ref{sec:axi} the special case of axisymmetric odd modes following \cite{Schnitzer:15plas}. In each section we compare our asymptotic results with an exact semi-analytical scheme based on separation of variables in bi-spherical coordinates. In \S\ref{sec:comparison} we compare our approximations for the eigenvalues with approximations obtained by others starting from the latter numerical scheme. Lastly, in \S\ref{sec:conc} we discuss how the present analysis could be generalised to related geometries, limits where our asymptotic results breakdown and an alternative analysis is desirable, and anticipate future applications of these results to excitation scenarios. 
\begin{figure}[t!]
\begin{center}
\includegraphics[trim={0 0 2cm 0},clip,scale=0.45]{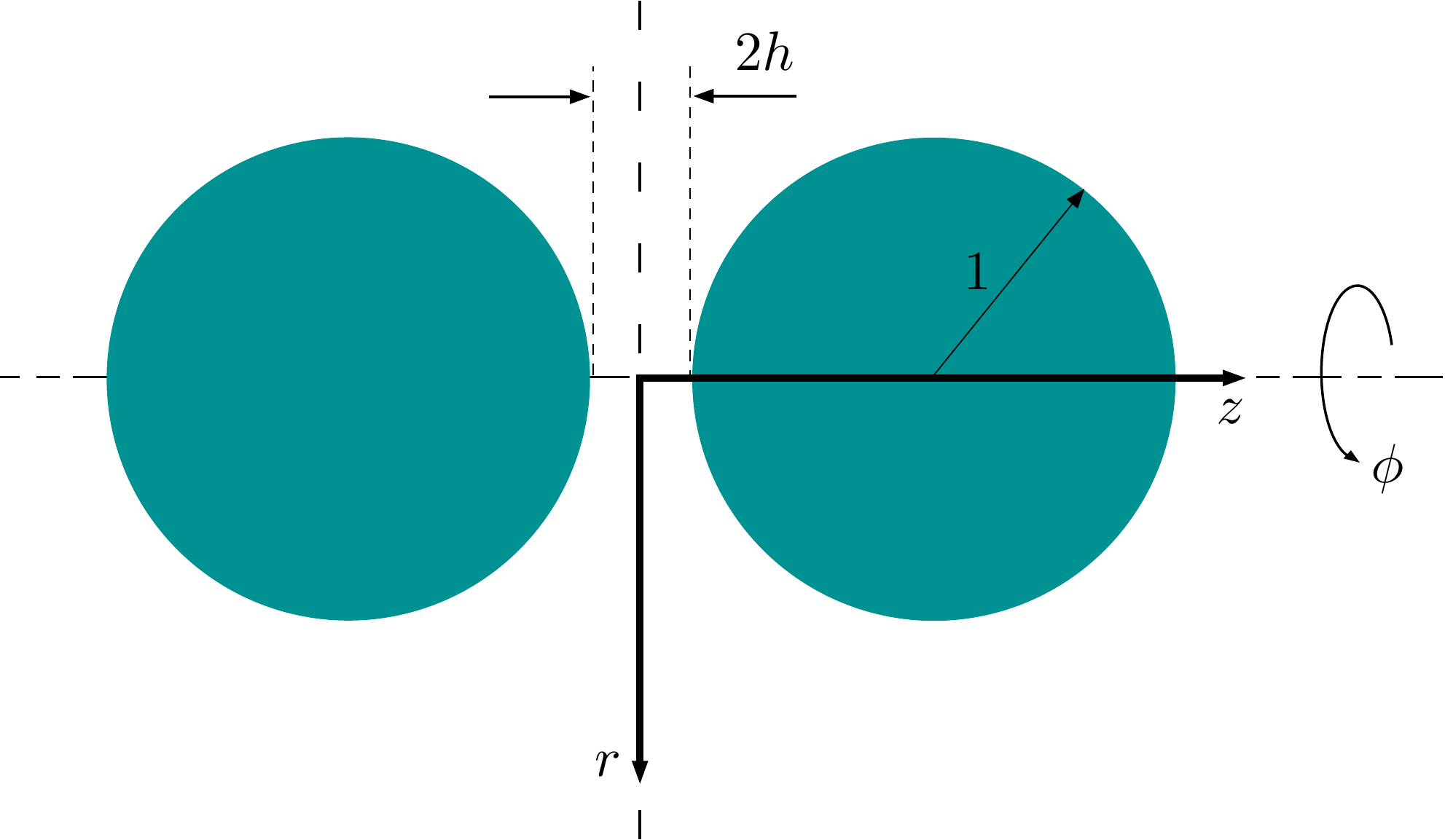}
\caption{Schematic of the scaled geometry.}
\label{fig:schematic}
\end{center}
\end{figure}

\section{Formulation of the problem}\label{sec:formulation}
\subsection{Plasmonic eigenvalue problem for a sphere dimer}\label{ssec:evproblem}
Our interest is in the plasmonic eigenvalues and eigenfunctions of a pair of identical homogeneous spheres surrounded by a homogenous background medium. The scale invariance of the plasmonic eigenvalue problem naturally lends itself to a dimensionless formulation where lengths are normalised by the radius of the spheres. The scaled geometry, shown in figure \ref{fig:schematic}, is characterised by a single parameter $h$ defined as the ratio of the gap width and the sphere diameter. Since we are not enforcing a certain normalisation of the eigenfunctions, the potentials possess a multiplicative freedom and we can consider them to be dimensionless. Furthermore, an immaterial additive freedom is eliminated by requiring the potentials to attenuate at large distances. It will be convenient to denote the potentials inside and outside the spherical inclusions by $\bar\varphi$ and $\varphi$, respectively. 

The problem consists of Laplace's equation inside the spherical inclusions,
\begin{equation}\label{laplace spheres}
\nabla^2\bar\varphi=0,
\end{equation}
and in the background medium, 
\begin{equation}\label{laplace back}
\nabla^2\varphi=0;
\end{equation}
continuity of potential,
\begin{equation}\label{cont}
\bar\varphi=\varphi,
\end{equation}
and of electric displacement, 
\begin{equation}\label{disp}
\epsilon\pd{\bar\varphi}{n}=\pd{\varphi}{n},
\end{equation}
on the spherical interfaces, where $\epsilon$ is the eigenvalue and $\partial/\partial{n}=\bn\bcdot\bnabla$, $\bn$ being the normal unit vector pointing into the background medium; and attenuation,
\begin{equation}\label{atten}
\varphi\to 0 \quad \text{as} \quad |\bx|\to\infty,
\end{equation} 
where $\bx$ is the position vector measured from the centre of the gap, say. 

\subsection{Symmetries}\label{ssec:sym}
Consider the cylindrical coordinates $(r,z,\phi)$ shown in figure \ref{fig:schematic}. 
Note that the geometry is symmetric about both the $z$ axis and the plane $z=0$ bisecting the gap. The axial symmetry implies that the eigenfunctions posses the form
\begin{equation}\label{azimuth}
\bar\varphi(r,z,\phi)=\bar\psi(r,z)\begin{cases}\cos(m\phi)\\ \sin(m\phi)\end{cases}, \quad \varphi(r,z,\phi)=\psi(r,z)\begin{cases}\cos(m\phi)\\ \sin(m\phi)\end{cases},
\end{equation}
where $m=0,1,2,\ldots$. 
Owing to the symmetry about the plane $z=0$, the modes can be further characterised as being either odd or even with respect to that plane. This allows us to consider only the half space $z>0$,  
with the odd modes satisfying 
\begin{equation}\label{symmetry odd}
\psi=0 \quad \text{at} \quad z=0
\end{equation}
and the even modes satisfying 
\begin{equation}\label{symmetry even}
\pd{\psi}{z}=0 \quad \text{at} \quad z=0.
\end{equation}

\subsection{Exact semi-analytical solutions}\label{ssec:exact}
The above plasmonic eigenvalue problem has been previously studied using separation of variables in bi-spherical coordinates \cite{Ruppin:82,Klimov:14} and using transformation optics followed by separation of variables in spherical coordinates \cite{Pendry:13}. In any case one finds an infinite-tridiagonal-matrix eigenvalue problem that in general must be truncated and solved numerically for the eigenvalues and eigenvectors, the latter being the coefficients in an infinite-series representation of the eigenpotentials. We have implemented this scheme for later comparison with our asymptotic results (see \cite{Klimov:14} for details). 

Computed eigenvalues corresponding to odd modes are shown in figures \ref{fig:odd_m1},\ref{fig:odd_m2} and \ref{fig:odd_m0} for $m=1,2$ and $0$, respectively.  Note that the eigenvalues of the odd modes lie below the accumulation point, i.e., $\epsilon<-1$, with $\epsilon\to-\infty$ as $h\to0$. Eigenvalues corresponding to even modes are shown in figures \ref{fig:even_m0}--\ref{fig:even_m2} for $m=0,1$ and $2$, respectively.
Note that there are in general two families of even modes, one with $\epsilon>-1$ and the other with $\epsilon<-1$. Even modes with $\epsilon>-1$ only exist for $h$ smaller than a critical value, which is different for each mode; these modes are similar to the even modes of a cylindrical dimer in that the eigenvalues satisfy $\epsilon\to0$ as $h\to0$, though in the latter case the modes exist for all $h>0$. Even modes with $\epsilon<-1$, termed ``anomalous'' in \cite{Pendry:13}, exist for all $h$ and in fact limit to the modes of an isolated sphere as $h\to\infty$. They are anomalous in that the eigenvalues tend to constants as $h\to0$, which has no analogue in the case of a cylindrical dimer. Moreover, the anomalous modes have their field excluded from the gap in the near-contact limit, whereas all the other modes of a sphere dimer have their field confined to the gap in that limit. Note that for each family of modes, and for each $m$, we use a second integer, $n=0,1,2,\ldots$, to enumerate the eigenvalues with increasing closeness to the accumulation point.

\section{The near-contact limit}\label{sec:limit}
\begin{figure}[t!]
\begin{center}
\includegraphics[trim={1cm 2cm 0 0},clip,scale=0.7]{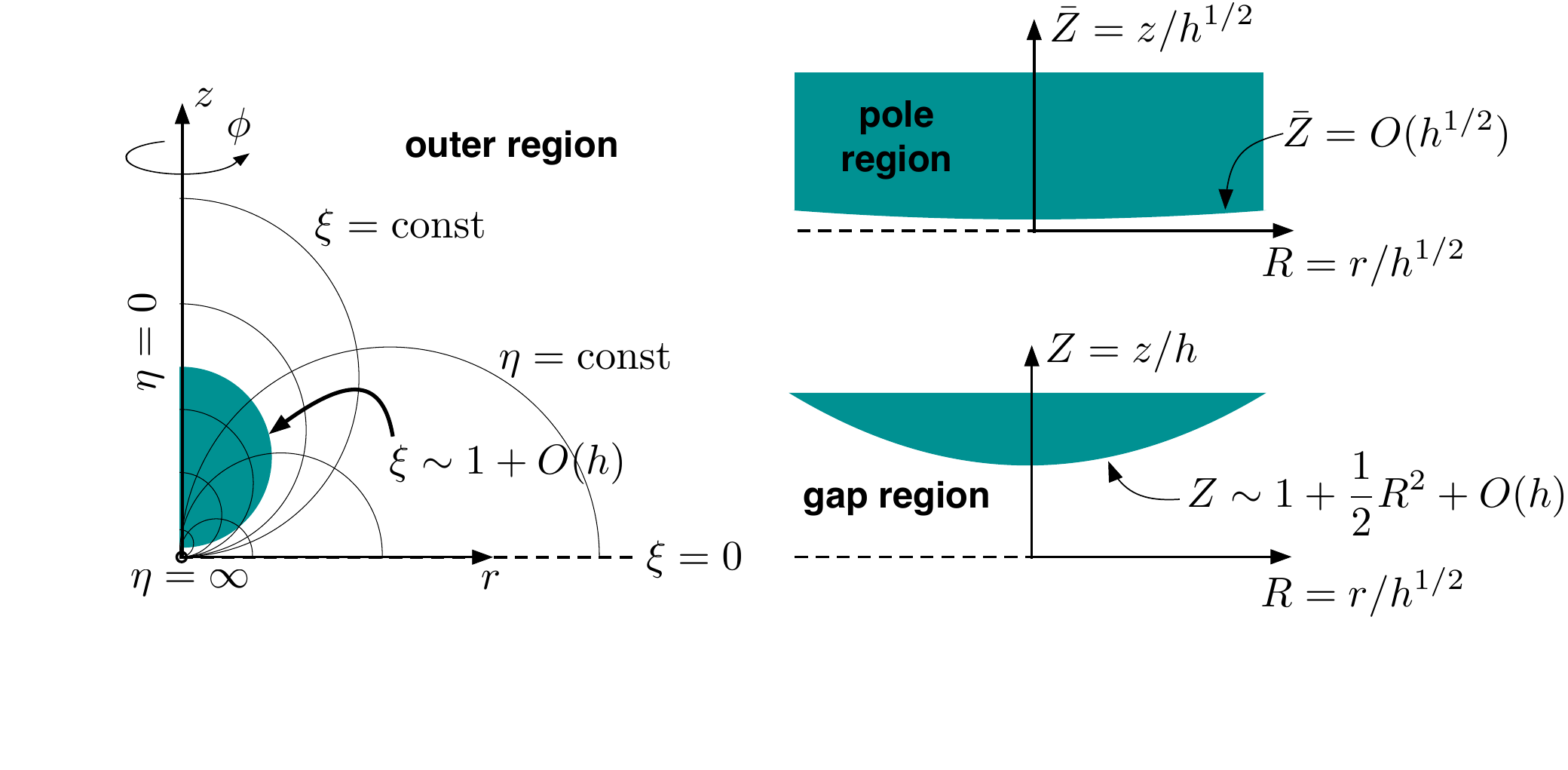}
\caption{The gap, pole and outer regions considered in the near-contact analysis.}
\label{fig:domains}
\end{center}
\end{figure}

In what follows our interest is in the limit $h\ll1$. Note that the limit is taken for a fixed arbitrary mode, which in particular implies $m,n=O(1)$. 
Since the near-contact limit represents a ratio between two geometric length scales, we anticipate the potentials to have spatially nonuniform asymptotics \cite{Van:pert}. We shall accordingly base our analysis on the method of matched asymptotic expansions \cite{Hinch:91}. In the present section we prepare for this by identifying three distinguished asymptotic regions (see figure \ref{fig:domains}).

\subsection{Outer region} \label{ssec:outer}
Consider the outer limit: $h\to0$ with $\bx$ fixed. The boundary of the sphere in the half space $z>0$ can be written as $F(r,z;h)=0$, where 
\begin{equation}\label{sphere}
F=[z-(1+h)]^2+r^2-1.
\end{equation} 
In the outer limit, \eqref{sphere} gives
\begin{equation}
F(r,z;h) \sim F_0(r,z) + O(h), \quad F_0(r,z) = (z-1)^2+r^2-1,
\end{equation}
where the nominal boundary $F_0(r,z)=0$ is a unit sphere tangent to the symmetry plane. 
In the outer region, the potentials $\bar\varphi$ and $\varphi$ satisfy \eqref{laplace spheres}--\eqref{atten}, with the interfacial conditions mapped to the nominal boundary by use of Taylor expansions. In general, the outer potentials may diverge as the origin is approached \cite{Jeffrey:78}. Indeed, the asymptotic behaviour in the latter limit is dictated by matching with inner regions describing the gap region and the adjacent region within the inclusion. 

We shall see that in most cases a detailed solution of the outer region is not required, at least at leading order. Otherwise, it will be convenient to employ tangent-sphere coordinates $(\xi,\eta,\phi)$ \cite{Moon:61}, where
\begin{equation}\label{tangent}
r = \frac{2\eta}{\xi^2+\eta^2}, \quad z = \frac{2\xi}{\xi^2+\eta^2},
\end{equation}
and $\phi$ is as before the azimuthal angle. As schematically shown in figure \ref{fig:domains}, the sphere boundary is $\xi=1+O(h)$, the symmetry plane is $\xi=0$, the origin is $\eta=\infty$ and infinity is approached in the limit $\xi^2+\eta^2\to0$.

\subsection{Gap region}  \label{ssec:gap}
Close to the gap the boundaries are approximately paraboloidal, hence the separation between the spheres  remains $O(h)$ over $O(h^{1/2})$ radial distances. Since the governing equations are linear and homogeneous, and in the absence of additional small parameters, the length scale on which the potential varies is determined by the geometry of the confining boundaries. Thus consider the inner-gap limit: $h\to0$ with the stretched co-ordinates 
\begin{equation}\label{gap scaling}
Z=z/h, \quad R=r/h^{1/2}
\end{equation}
fixed. In terms of the gap coordinates \eqref{gap scaling}, the sphere boundary follows from \eqref{sphere} as 
\begin{equation}\label{gap boundary}
Z = H(R;h)\sim H_0(R) + O(h), \quad H_0(R)=1+\frac{1}{2}R^2.
\end{equation}
The gap potential, $\Phi(R,Z)=\psi(r,z)$, satisfies Laplace's equation \eqref{laplace back} in the form
\begin{equation}\label{gap laplace}
h^{-1}\pd{^2\Phi}{Z^2}+\frac{1}{R}\pd{}{R}\left(R\pd{\Phi}{R}\right)-\frac{m^2}{R^2}\Phi=0,
\end{equation}
for $0<Z<H(R;h)$. 
Given \eqref{symmetry odd} and \eqref{symmetry even}, the odd and even modes respectively satisfy the symmetry conditions 
\begin{equation}\label{sym gap odd}
\Phi = 0 \quad \text{at} \quad Z=0
\end{equation}
and
\begin{equation}\label{sym gap even}
\pd{\Phi}{Z}=0 \quad \text{at} \quad Z=0.
\end{equation}
The interfacial conditions satisfied by $\Phi$ at $Z=H(R;h)$ will be provided in the next subsection. Furthermore, $\Phi$ is required to asymptotically match with the outer region in the limit $R\to\infty$. 

\subsection{Pole region} \label{ssec:pole}
Consider now the region within the spherical inclusion that is adjacent to the gap. The scaling of the gap region, together with the continuity condition \eqref{cont}, suggests that $\bar\psi$ varies over $O(h^{1/2})$ distances along the interface. On such a small scale, the spherical domain is effectively unbounded in the transverse direction. Thus, the linearity and symmetry of Laplace's equation suggest that $\bar\psi$ varies over comparably short distances in that direction. We accordingly introduce the inner-pole limit: $h\to0$ with the stretched coordinates 
\begin{equation}\label{pole}
\bar{Z}=z/h^{1/2}, \quad R=r/h^{1/2}
\end{equation}
fixed. In terms of the pole coordinates \eqref{pole}, the sphere boundary reads as
\begin{equation}\label{pole interface}
\bar{Z}=h^{1/2}{H}(R;h).
\end{equation}
Note that since $H=O(1)$, the boundary is approximately flat in the pole limit. The pole potential, $\bar\Phi(R,\bar Z)=\bar\psi(r,z)$, satisfies Laplace's equation \eqref{laplace spheres} in the form
\begin{equation}\label{lap pole}
\pd{^2\bar\Phi}{\bar Z^2}+\frac{1}{R}\pd{}{R}\left(R\pd{\bar\Phi}{R}\right)-\frac{m^2}{R^2}\bar\Phi=0,
\end{equation}
for $\bar{Z}>h^{1/2}H(R;h)$. 
The continuity condition \eqref{cont} now reads as
\begin{equation}\label{cont pole}
\bar\Phi=\Phi \quad \text{on} \quad Z=H(R;h), \quad \bar{Z}=h^{1/2}{H}(R;h),
\end{equation}
while the displacement-continuity condition \eqref{disp} reads as
\begin{multline}\label{disp pole}
\epsilon h^{1/2}\left(\pd{\bar\Phi}{\bar{Z}}-h^{1/2}\frac{\dd{H}}{\dd R}\pd{\bar\Phi}{R}\right)
\\ =\pd{\Phi}{Z}-h\frac{\dd H}{\dd R}\pd{\Phi}{R}
 \quad \text{at} \quad Z=H(R;h), \quad \bar{Z}=h^{1/2}{H}(R;h).
\end{multline}
Lastly, $\bar{\Phi}$ must asymptotically match with the outer region in the limit $R^2+\bar Z^2\to\infty$.

\section{Odd modes}\label{sec:odd}
\subsection{Gap region}
We may assume without loss of generality that the potential in the gap is $O(1)$. We accordingly write
\begin{equation}
\Phi(R,Z) = \Phi_0(R,Z)+o(1),
\end{equation}
where the magnitude of the error term is discussed in \S\S\ref{ssec:comparisonodd}.
From \eqref{gap laplace}, $\Phi_0$ satisfies
\begin{equation}
\pd{^2\Phi_0}{Z^2}=0,
\end{equation}
while \eqref{sym gap odd} gives the symmetry condition
\begin{equation}
\Phi_0=0\quad\text{at} \quad Z=0.
\end{equation}
It follows that
\begin{equation}\label{odd gap potential}
\Phi_0 = A(R) Z,
\end{equation}
where $A(R)$ is an unknown radial distribution. We see that, in the near-contact limit, the odd modes are characterised by a transverse electric field in the gap that varies radially (and azimuthally if $m\ne0$) on a scale large compared with the gap width. 

\subsection{Pole potential and eigenvalue scaling} 
Consider next the pole region. The continuity condition \eqref{cont pole} suggests that the magnitude of the pole potential is comparable to that of the gap potential. We accordingly assume the expansion
\begin{equation}
\bar\Phi(R,\bar Z)=\bar\Phi_0(R,\bar Z)+ o(1). 
\end{equation}
The eigenvalue scaling then follows by inspection of the displacement condition \eqref{disp pole}:
\begin{equation}\label{eps scaling odd order}
\epsilon= O(h^{-1/2}).
\end{equation}

\subsection{Effective eigenvalue problem}
We expand the eigenvalue as 
\begin{equation}\label{eps scaling odd}
\epsilon= - \alpha h^{-1/2} + o(h^{-1/2}),
\end{equation}
where $\alpha$ is a positive constant to be determined from an effective eigenvalue problem governing the leading-order pole potential $\bar\Phi_0$. 
From \eqref{lap pole}, $\bar\Phi_0$ satisfies Laplace's equation in the half space $\bar{Z}>0$,
 \begin{equation}\label{lap pole 0}
\pd{^2\bar\Phi_0}{\bar Z^2}+\frac{1}{R}\pd{}{R}\left(R\pd{\bar\Phi_0}{R}\right)-\frac{m^2}{R^2}\bar\Phi_0=0,
\end{equation}
while the interfacial conditions \eqref{cont pole} and \eqref{disp pole}, in conjunction with \eqref{eps scaling odd}, give
\begin{equation}\label{pole condition 1}
\bar\Phi_0=A(R)H_0(R) \quad \text{at} \quad \bar{Z}=0
\end{equation}
and
\begin{equation}\label{pole condition 2}
-\alpha \pd{\bar\Phi_0}{\bar{Z}}=A(R) \quad \text{at} \quad \bar{Z}=0.
\end{equation}
Eliminating $A(R)$ we obtain
\begin{equation}\label{bc odd}
\bar\Phi_0+\alpha H_0\pd{\bar\Phi_0}{\bar{Z}}=0 \quad \text{at} \quad \bar{Z}=0,
\end{equation}
a self-contained robin-type condition with non-constant coefficients for $\bar\Phi_0$. 

To close the problem governing $\bar\Phi_0$ a condition on its behaviour in the limit $R^2+\bar{Z}^2\to\infty$ should in principle be deduced from asymptotic matching with the outer region. It is more expedient to assume, subject to verification by matching, that the outer potential within the sphere is asymptotically small compared to the potential in the gap and pole regions. In that case, matching implies the attenuation condition 
\begin{equation}\label{odd matching decay}
\bar\Phi_0\to0 \quad \text{as} \quad R^2+\bar{Z}^2\to\infty.
\end{equation}
Since $\bar\Phi_0$ satisfies Laplace's equation, \eqref{odd matching decay} can be refined to 
\begin{equation}\label{odd matching estimate}
\bar\Phi_0 = O\left\{\left(R^2+\bar{Z}^2\right)^{-\frac{1+m}{2}}\right\} \quad \text{as} \quad R^2+\bar{Z}^2\to\infty.
\end{equation}
Eqs.~\eqref{lap pole 0}, \eqref{bc odd} and \eqref{odd matching decay} then constitute an effective eigenvalue problem governing the leading-order pole potential $\bar\Phi_0$ and scaled eigenvalue $\alpha$. Once $\bar\Phi_0$ is determined, the gap potential can be found using \eqref{pole condition 1}. 

\subsection{Solution of the effective eigenvalue problem}
We look for solutions in the form
\begin{equation}\label{hankel odd}
\bar{\Phi}_0(R,\bar{Z})=\underset{s\to R}{\mathcal{H}_m}\hat{\bar\Phi}(s,\bar{Z}), 
\end{equation}
where
\begin{equation}
\underset{s\to R}{\mathcal{H}_m}f(s)=\int_0^{\infty} f(s)J_m(Rs)s\,ds
\end{equation}
is the Hankel transform of order $m$ and $J_m$ is the usual Bessel function of order $m$ \cite{Sneddon:Book}. Note that the inverse Hankel transform has the same form as the forward transform. Substitution of \eqref{hankel odd} into Laplace's equation \eqref{lap pole 0} and using the attenuation condition \eqref{odd matching decay} yields
\begin{equation}\label{Y def}
\hat{\bar{\Phi}}(s,\bar{Z})=\frac{1}{s}Y(s)e^{-s\bar{Z}},
\end{equation}
where $Y(s)$ is an unknown function of the transform variable $s$. In terms of this function, the boundary condition \eqref{bc odd} reads as
\begin{equation}\label{Y step 1}
\underset{s\to R}{\mathcal{H}_m}\left[\left(\frac{1}{s}-\alpha\right)Y(s)\right]-\frac{\alpha}{2}R^2\underset{s\to R}{\mathcal{H}_m}Y(s)=0.
\end{equation}
Taking the inverse transform of \eqref{Y step 1}, using the identity 
\begin{equation}\label{Identity} 
-R^2 \underset{s\to R}{\mathcal{H}_m}f(s)=\underset{s\to R}{\mathcal{H}_m}\left[\frac{1}{s}\pd{}{s}\left(s\pd{{f}}{s}\right)-\frac{m^2}{s^2}{f}\right],
\end{equation}
which is valid if the transform exists and 
\begin{equation}\label{cond1}
s\frac{\dd f}{\dd s}=o(s^{-m}) \quad \text{as} \quad s\to0
\end{equation} 
(see \cite{Sneddon:Book}), yields 
\begin{equation}\label{Y eq odd}
\frac{\dd^2Y}{\dd s^2}+\frac{1}{s}\frac{\dd Y}{\dd s}+\left(\frac{2}{\alpha s}-2-\frac{m^2}{s^2}\right)Y=0.
\end{equation}

Analysis of \eqref{Y eq odd} for large $s$ \cite{Bender:13} suggests the transformation
\begin{equation}
Y(s) = s^m e^{-\sqrt{2}s}T(p), \quad p = 2\sqrt{2}s,
\end{equation}
where $T(p)$ satisfies the associated Laguerre equation \cite{Abramowitz:book}:
\begin{equation}\label{assoc Lag odd}
p\frac{\dd^2T}{\dd p^2}+(1+\nu-p)\frac{\dd T}{\dd p}+nT=0,
\end{equation}
with parameters
\begin{equation}\label{param n}
\nu=2m, \quad n = \frac{\sqrt{2}-(1+2m)\alpha}{2\alpha}. 
\end{equation}

For non-negative integer $n$, one solution of  \eqref{assoc Lag odd} is the polynomial  
\begin{equation}
L_n\ub{\nu}(x)=\frac{(\nu+1)_n}{n!} {_1F_1}(-n;\nu+1;x),
\end{equation}
where $(t)_n=\Gamma(t+n)/\Gamma(t)$ is  the Pochhammer symbol and ${_1F_1}$ the hypergeometric function, e.g.,
\begin{multline}
L_0\ub{\nu}(x)=1, \quad L_1\ub{\nu}(x)=-x+\nu+1, \\ L_2\ub{\nu}(x)=\frac{1}{2}\left[x^2-2(\nu+2)x+(\nu+1)(\nu+2)\right];
\end{multline}
if $\nu$ is also an integer, as in the present case where $\nu=2m$, these polynomials are termed associated Laguerre polynomials, otherwise they are called the generalised Laguerre function \cite{Abramowitz:book}.
It can be shown that the second independent solution for non-negative integer $n$, as well as all solutions for other $n$ values, are either too singular as $p\to0$ to satisfy \eqref{cond1} [and the attenuation rate \eqref{odd matching estimate}] or grow too fast as $p\to\infty$ for the transform \eqref{hankel odd} to exist. Thus $n$ must be a non-negative integer whereby the scaled eigenvalues follow from \eqref{param n} as
\begin{equation}\label{odd ev m pos}
\alpha = \frac{\sqrt{2}}{1+2n+2m}, \quad n=0,1,2,\ldots, \quad m=1,2,3,\ldots,
\end{equation}
with corresponding eigenfunctions
\begin{equation}\label{odd nonaxi Y}
Y(s) = s^{m}e^{-\sqrt{2}s}L_n\ub{2m}(2\sqrt{2}s), \quad n=0,1,2,\ldots, \quad m=1,2,3,\ldots,
\end{equation}
where we have set the arbitrary multiplicative constants to unity. 
\begin{figure}[t!]
\begin{center}
\includegraphics[scale=0.4]{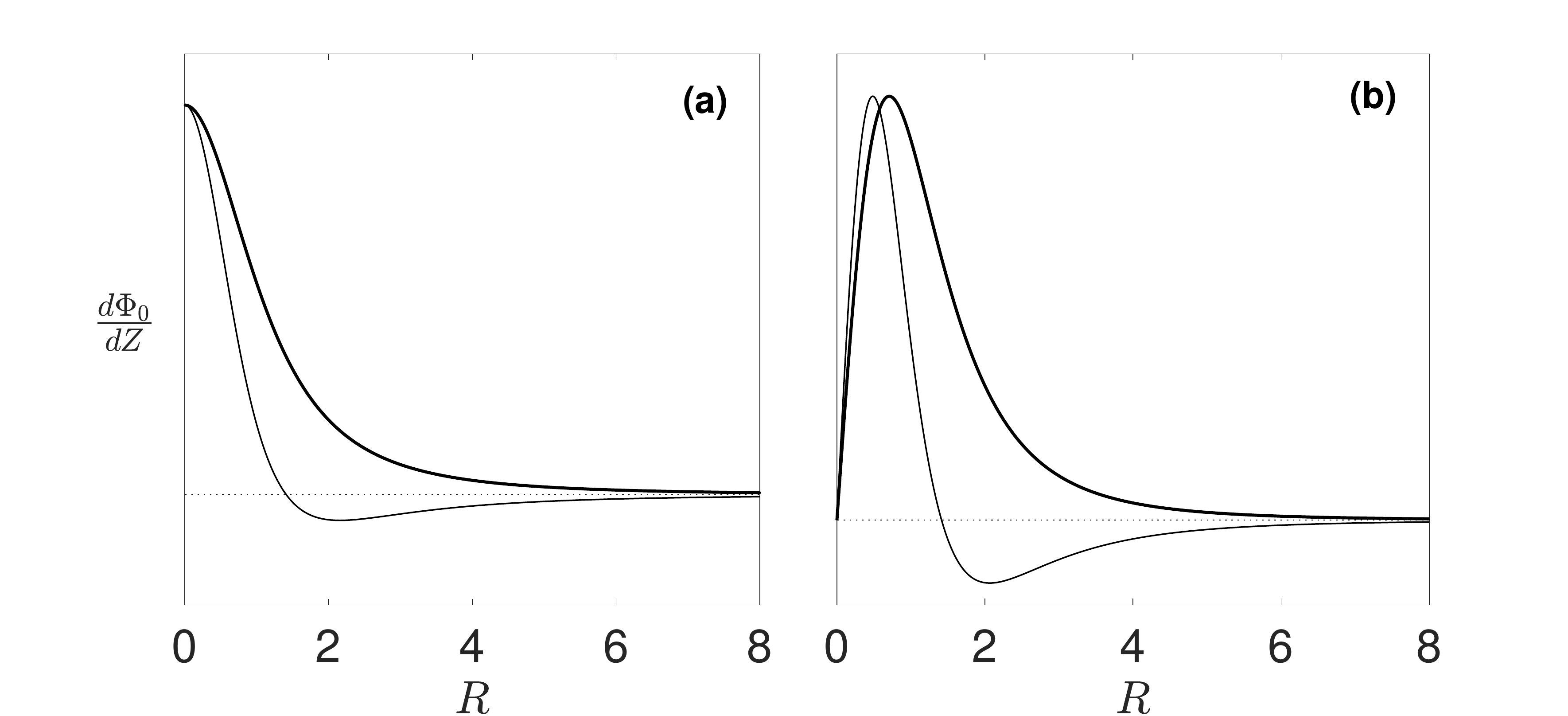}
\caption{Radial distributions \eqref{odd fields m0} and \eqref{odd fields m1} of the transverse field in the gap for odd modes with $m=0$ (a) and $m=1$ (b). Thick and thin lines depict modes $n=0$ and $n=1$, respectively. In the case $m=0$ we shall derive more accurate asymptotics in \S\ref{sec:axi}.}
\label{fig:oddnonaxi_fields}
\end{center}
\end{figure}

\subsection{Eigenfunctions in physical space}
The eigenfunctions \eqref{odd nonaxi Y} can be inverted to give, for example, the radial distribution of the transverse field in the gap [cf.~\eqref{odd gap potential} and \eqref{pole condition 1}]:
\begin{equation}\label{odd inverse}
\frac{\dd \Phi_0}{\dd Z}=\frac{1}{H_0(R)}\int_0^{\infty}s^{m}e^{-\sqrt{2}s}L_n\ub{2m}(2\sqrt{2}s)J_m(Rs)\,\dd s.
\end{equation}
It can be shown from \eqref{odd inverse} that $\dd\Phi_0/\dd Z=O(R^{-(m+3)})$ as $R\to\infty$. 
It follows that in that limit $\bar\Phi_0(R,0)=O(1/R^{1+m})$, as anticipated in \eqref{odd matching estimate}, while  $\Phi_0= O(Z/R^{3+m})$.
The quadrature \eqref{odd inverse} can be evaluated for given $m$ and $n$. In particular, for $(m,n)=(0,\{0,1\})$: 
\begin{equation}\label{odd fields m0}
\frac{\dd \Phi_0}{\dd Z} =\frac{1}{H_0(R)}\left\{\frac{1}{\left(2+R^2\right)^{1/2}},\frac{R^2-2}{\left(2+R^2\right)^{3/2}}\right\}.
\end{equation}
Similarly, 
\begin{equation}\label{odd fields m1}
\frac{\dd \Phi_0}{\dd Z}=\frac{1}{H_0(R)}\left\{\frac{R}{(2+R^2)^{3/2}},\frac{3R(R^2-2)}{(2+R^2)^{5/2}}\right\}
\end{equation}
and
\begin{equation}\label{odd fields m2}
\frac{\dd \Phi_0}{\dd Z}=\frac{1}{H_0(R)}\left\{\frac{3R^2}{(2+R^2)^{5/2}},\frac{15R^2(R^2-2)}{(2+R^2)^{7/2}}\right\}
\end{equation}
for $(m,n)=(1,\{0,1\})$ and $(m,n)=(2,\{0,1\})$, respectively.  We plot the transverse gap fields \eqref{odd fields m0} and \eqref{odd fields m1} in figure \ref{fig:oddnonaxi_fields}.

\begin{figure}[b!]
\begin{center}
\includegraphics[scale=0.4]{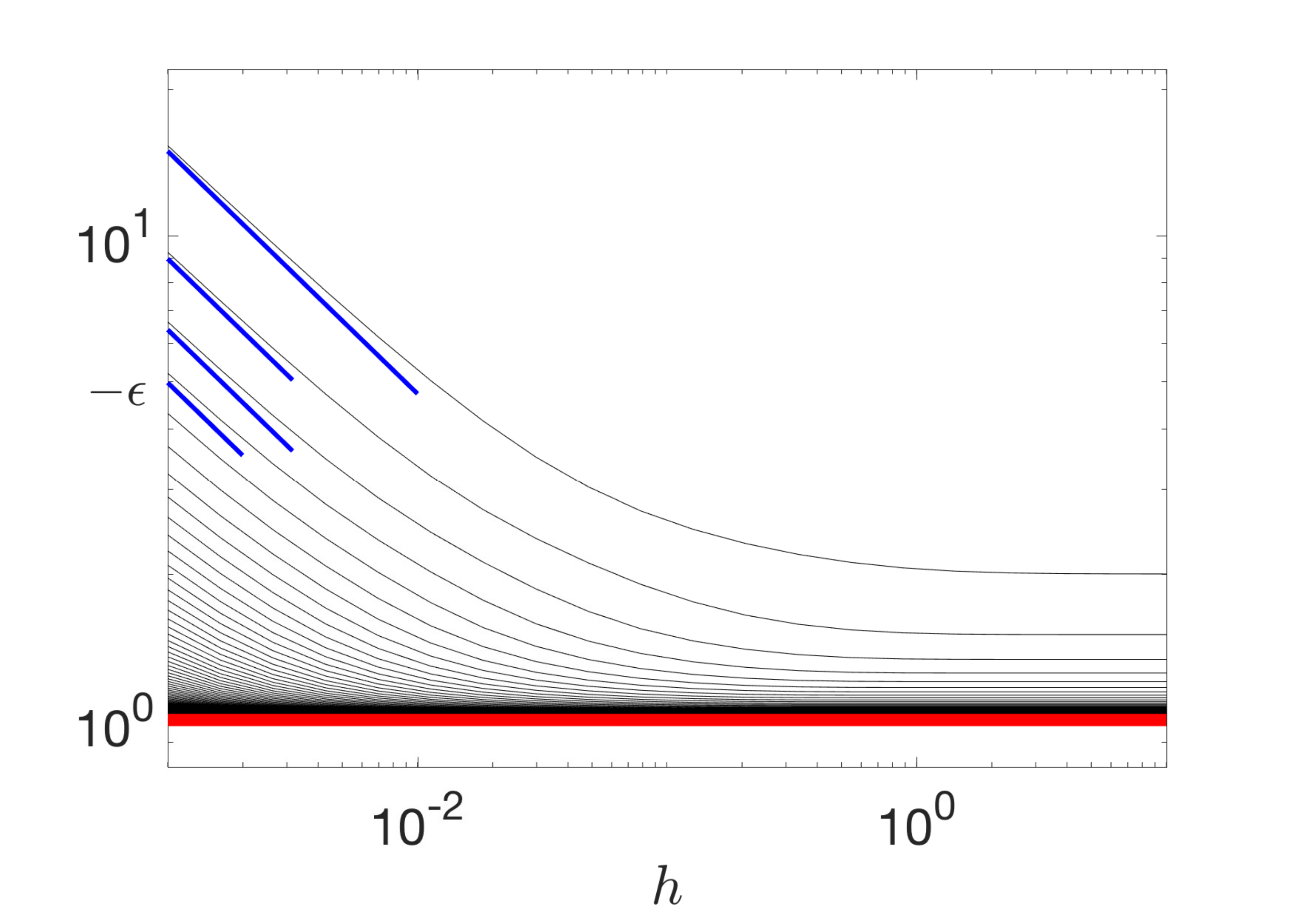}
\caption{Eigenvalues corresponding to non-axisymmetric ($m=1$) odd modes as a function of $h$, half the dimensionless gap width. The thick blue lines are the asymptotic predictions \eqref{eps scaling odd}, with $\alpha$ given by \eqref{odd ev m pos} for $n=0,1,2,3$. The thin black  lines are exact values obtained from the semi-numerical scheme described in \S\S\ref{ssec:exact}. The thick red line marks the accumulation point $\epsilon=-1$.}
\label{fig:odd_m1}
\end{center}
\end{figure}
\begin{figure}[t!]
\begin{center}
\includegraphics[scale=0.4]{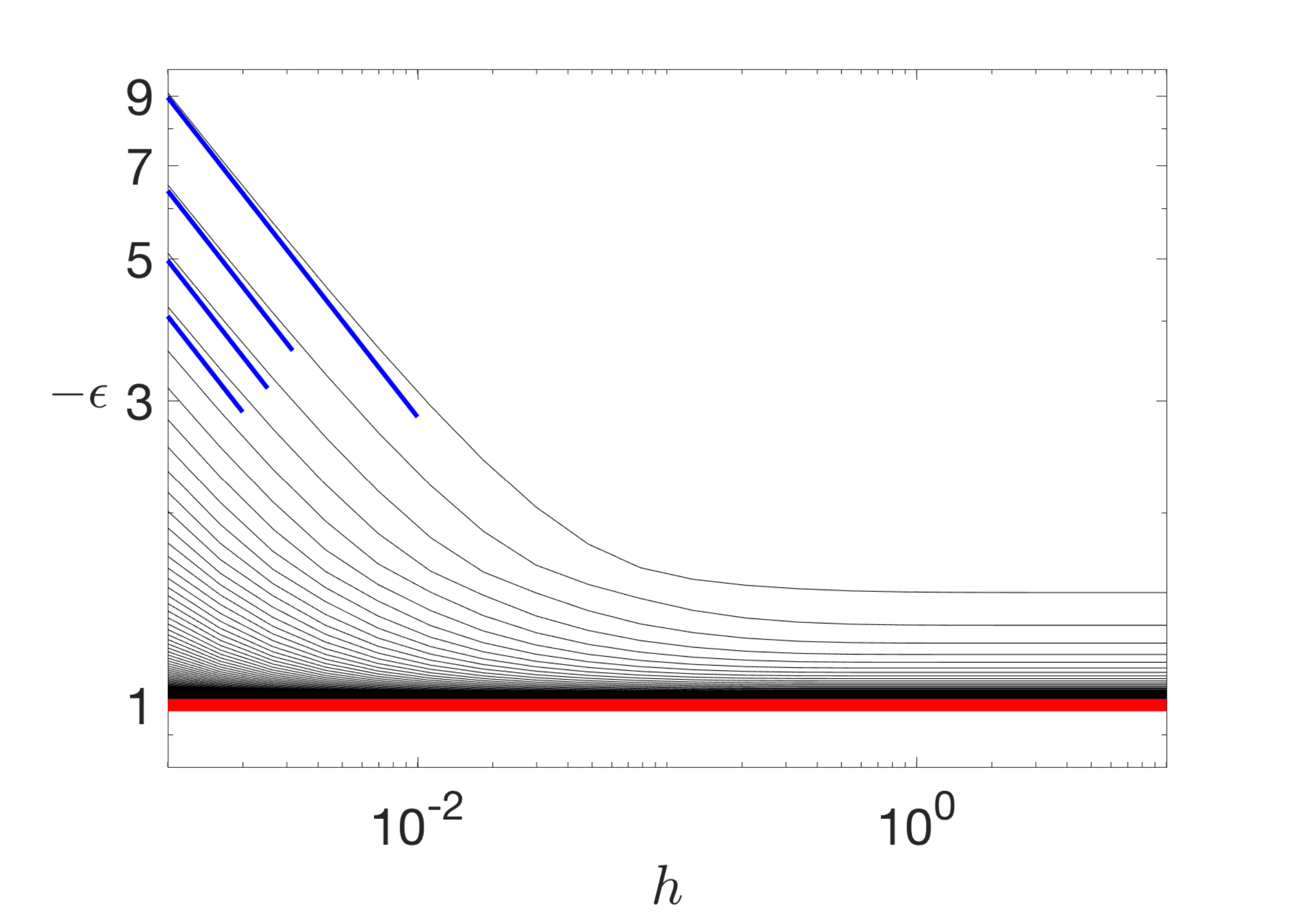}
\caption{Same as figure \ref{fig:odd_m1} but for non-axisymmetric odd modes with $m=2$.}
\label{fig:odd_m2}
\end{center}
\end{figure}

\subsection{Comparison with exact semi-analytical solutions}\label{ssec:comparisonodd}
Figures \ref{fig:odd_m1},\ref{fig:odd_m2} present, respectively for $m=1,2$, a comparison between the  asymptotic prediction \eqref{eps scaling odd}, with $\alpha$ given by \eqref{odd ev m pos}, and the eigenvalues computed using the semi-analytical scheme discussed in \S\S\ref{ssec:exact}. The agreement is excellent for small $h$, though for reasons discussed in \S\ref{sec:conc} the agreement is delayed to smaller $h$ as $n$ increases. In contrast, for $m=0$ the computed eigenvalues are found to converge extremely slowly to the asymptotic result, suggesting that the error in the axisymmetric case is relatively large. 
 
From inspection of the gap-pole equations, one might expect a relative asymptotic error of $O(h^{1/2})$. Recall, however, that the analysis in this section is predicated on the assumption that $\bar\varphi$, the outer potential within the sphere, is asymptotically small compared to the potential in the pole and gap regions and accordingly that $\bar\Phi_0$ satisfies the attenuation condition \eqref{odd matching decay}. By analysing the leading-order outer problem, it can be shown that for $m\ne0$ our naive leading-order solutions are in fact algebraically accurate and that the outer potentials are $\bar\varphi,\varphi=O(h^{(m+1)/2})$. For $m=0$, however, we shall see that $\bar\varphi$ is approximately uniform and only $O(1/\ln h)$ smaller than the pole potential. The matching condition \eqref{odd matching decay} must accordingly be modified already at that order, leading to logarithmic corrections that are practically important. 

\section{Axisymmetric odd modes}\label{sec:axi}
Our goal in this section is to develop an accurate asymptotic description of the axisymmetric modes. We shall closely follow our analysis in \cite{Schnitzer:15plas} of this case.
\subsection{Gap, pole and internal outer regions}
We start by generalising expansion \eqref{eps scaling odd}:
\begin{equation}\label{axi epsi expansion}
\epsilon \approx - \alpha(\ln h) h^{-1/2},
\end{equation}
where the approximation sign will be used to denote an error which is algebraic, i.e., scaling with some power of $h$. In \eqref{axi epsi expansion}, and throughout the analysis in this section, we collect together terms that are asymptotically separated by powers of $\ln h$. The leading-order pole and gap potentials are still denoted by $\bar\Phi_0$ and $\Phi_0=A(R)Z$, respectively, only that by leading order we now mean that the relative error is algebraically small. Hence we allow $\bar\Phi_0$ and $A(R)$ to depend upon $\ln h$. With these conventions, Eqs. \eqref{lap pole 0}--\eqref{bc odd} governing $\bar\Phi_0$ remain in the same form, whereas the attenuation condition \eqref{odd matching decay} requires modification. 

To see this, consider the outer potential within the sphere $\bar\varphi$. The displacement condition \eqref{disp} in conjunction with the largeness of $|\epsilon|$ implies that $\bar\varphi$ approximately satisfies a homogeneous Neumann condition on the tangent-sphere boundary. Accordingly, subject to matching with the pole region (in the limit where the origin is approached), we stipulate a uniform leading-order solution, say
\begin{equation}\label{outer inside odd axi}
{\bar\varphi} \approx V,
\end{equation}
where $V$ is a constant. This, in turn, implies through matching that the pole potential satisfies 
\begin{equation}\label{axi pole matching}
\bar\Phi_0\to V \quad \text{as} \quad R^2+\bar{Z}^2\to\infty
\end{equation}
instead of the attenuation condition \eqref{odd matching decay}. From \eqref{pole condition 1}, the gap potential accordingly satisfies
 \begin{equation}\label{axi gap behaviour}
 \Phi_0(R,Z) \sim \frac{2V}{R^2}Z \quad \text{as} \quad R\to\infty.
 \end{equation}
(Note that \eqref{outer inside odd axi} is impossible for $m\ne0$ and the leading-order outer potential is instead forced by matching with the attenuating pole solutions we found in \S\ref{sec:odd}. Conversely, the latter situation leads to a contradiction for $m=0$, since in that case the attenuating pole solution produces a net flux into the sphere, which is incompatible with the Neumann condition in the outer region. Hence the uniform solution \eqref{outer inside odd axi} holds with $V=o(1)$ yet large compared with $O(h^{1/2})$.)

\subsection{External outer region}
Outside the sphere we expand the outer potential as 
\begin{equation}
\varphi\approx\varphi_0(\xi,\eta),
\end{equation} 
where $(\xi,\eta)$ are the tangent-sphere coordinates defined in \S\S\ref{ssec:outer}. The potential $\varphi_0(\xi,\eta)$ satisfies Laplace's equation for $0<\xi<1$ and $0<\eta<\infty$; the Dirichlet boundary condition
\begin{equation}
\varphi_0=V \quad \text{at} \quad \xi=1,
\end{equation}
which follows from continuity with the internal outer potential \eqref{outer inside odd axi}; 
the symmetry condition 
\begin{equation}
\varphi_0=0 \quad \text{at} \quad \xi=0;
\end{equation}
attenuation at infinity 
\begin{equation}
\varphi_0\to0 \quad \text{as} \quad \eta\to\infty;
\end{equation}
as well as matching with the gap potential. Given \eqref{axi gap behaviour}, the latter requirement implies
\begin{equation}
\varphi_0\sim V\xi \quad \text{as} \quad \eta\to\infty.
\end{equation}

Thus, the outer potential is the same as if the spheres were perfectly conducting and held at opposite potentials $\pm V$. The requisite solution is known to be \cite{Jeffrey:78}
\begin{equation}\label{outer sol}
\varphi_0(\xi,\eta) = V(\xi^2+\eta^2)^{1/2}\underset{s\to \eta}{\mathcal{H}_0}\left(\frac{e^{-s}\sinh(s\xi)}{s\sinh s}\right).
\end{equation}
Note that at large distances from the spheres,
\begin{equation}
\varphi_0 \sim V\frac{\pi^2}{3}\frac{z}{(r^2+z^2)^{3/2}} \quad \text{as} \quad r^2+z^2\to\infty,  
\end{equation}
which relates the dipole moment of the axisymmetric odd modes and the voltage. 

\subsection{Charge balance}
In the electrostatic analogy of the outer solution, where a voltage $2V$ is applied between a pair of perfectly conducting spheres, the spheres are oppositely charged. As we shall see, attempting to calculate the charge using the outer solution \eqref{outer sol} gives a diverging result. Indeed, the overlap with the gap region must be accounted for and the capacitance is accordingly found to be logarithmically large in $h$ \cite{Jeffrey:78}.

In contrast, in the plasmonic case it is obvious on physical grounds that the net polarisation charge on each sphere must vanish. Indeed, the problem formulation implies the integral constraint
\begin{equation}\label{charge integral}
\oint \bn\bcdot\bnabla\varphi=0,
\end{equation}
where the integral is over the surface of the sphere in $z>0$, say. As the outer solution is the same as in the electrostatic problem, we conclude that the \emph{excess} polarisation charge in the gap must exactly balance the virtual net charge the sphere would have if held at a strictly uniform potential $V$ (and $-V$ on the sphere in $z<0$).

The above intuitive reasoning can be made formal. We split the integral in \eqref{charge integral} at some $1\ll\eta_0\ll h^{-1/2}$, which corresponds to an inner radial coordinate $h^{1/2}R_0=2/\eta_0+O(1/\eta_0^3)$. Using the gap expansion for $R<R_0$ and the outer expansion for $\eta>\eta_0$, \eqref{charge integral} yields 
\begin{equation}\label{split integral}
\int_0^{R_0}A(R)R\,\dd R+\int_0^{\eta_0}\left.\pd{\varphi_0}{\xi}\right|_{\xi=1}\frac{2\eta}{1+\eta^2}\,\dd \eta\approx 0.
\end{equation}
Since $A(R)=O(1/R^2)$ as $R\to\infty$ [cf.~\eqref{axi gap behaviour}], the first integral in \eqref{split integral}, which represents the leading contribution of the gap region, does not converge as $R_0\to\infty$. Indeed, using \eqref{pole condition 1} we find
\begin{equation}\label{gap contribution}
\int_0^{R_0}A(R)R\,\dd R \sim V\ln \frac{R_0^2}{2} + \int_0^{\infty}\frac{\bar{\Phi}(R,0)-V}{H_0(R)}R\,\dd R +o(1) \quad \text{as} \quad  R_0\to\infty,
\end{equation}
where the integral on the right hand side now clearly converges and represents the excess gap charge. Next, using the outer solution \eqref{outer sol} it can be shown that the outer contribution is
\begin{equation}\label{outer contribution}
\int_0^{\eta_0}\left.\pd{\varphi_0}{\xi}\right|_{\xi=1}\frac{2\eta}{1+\eta^2}\,\dd \eta\sim 2V(\ln\eta_0+\gamma)+o(1) \quad \text{as} \quad \eta_0\to\infty,
\end{equation}
where $\gamma=0.5772\ldots$ is the Euler-Gamma constant. Finally, adding \eqref{gap contribution} and \eqref{outer contribution}, the singular terms involving the arbitrary values $R_0$ and $\eta_0$ cancel out. We thereby derive the integral constraint
\begin{equation}\label{integral original}
V=-\frac{1}{\ln(2/h)+2\gamma}\int_0^{\infty}\frac{\bar{\Phi}(R,0)-V}{H_0(R)}\,R\,\dd R.
\end{equation}

\subsection{Integral-differential eigenvalue problem}
Consider now the pole problem in terms of the modified potential
\begin{equation}
\chi(R,\bar{Z})=\bar\Phi_0/V-1.
\end{equation}
Laplace's equation \eqref{lap pole 0} reads as 
 \begin{equation}\label{chi lap}
\pd{^2\chi}{\bar Z^2}+\frac{1}{R}\pd{}{R}\left(R\pd{\chi}{R}\right)=0,
\end{equation}
the boundary condition \eqref{bc odd} reads as
\begin{equation}\label{chi bc}
\chi+\alpha H_0\pd{\chi}{\bar{Z}}=-1 \quad \text{at} \quad \bar{Z}=0
\end{equation}
and the matching condition \eqref{axi pole matching} becomes the attenuation condition
\begin{equation}\label{chi decay}
\chi\to0 \quad \text{as} \quad R^2+\bar{Z}^2\to\infty. 
\end{equation}
In addition $\chi$ must satisfy the integral condition \eqref{integral original}, which becomes
\begin{equation}\label{chi integral}
\ln(2/h)+2\gamma+\int_0^{\infty}\frac{\chi(R,0)}{H_0(R)}\,R\,\dd R=0.
\end{equation}

\subsection{Infinite expansion in inverse logarithmic powers}
The above effective eigenvalue problem depends logarithmically upon $h$ via the integral constraint \eqref{chi integral}. Accordingly, it can be solved perturbatively in inverse logarithmic powers. In particular, the eigenvalue expansion is found as  \cite{Schnitzer:15plas}
\begin{equation}\label{log series}
\alpha \sim  \frac{\sqrt{2}}{1+2n}\left(1-\frac{4}{2n+1}\frac{1}{\ln({1}/{h})}+\cdots\right), \quad n=0,1,2,\ldots
\end{equation}
The leading order of \eqref{log series} agrees with the approximation obtained in \S\ref{sec:odd} in the case $m=0$. We next proceed to solve the effective eigenvalue problem exactly, thus recovering all the terms in the logarithmic expansion in one go. 

\subsection{Algebraically accurate solution}
In \cite{Schnitzer:15plas} $\chi$ was sought in terms of a Hankel transform. A Hankel transform is nothing but the two-dimensional Fourier transform of a radially symmetric function. We shall find the latter interpretation clearer when transforming the boundary condition \eqref{chi bc}, as the constant term requires carrying out the transform in the sense of distributions \cite{Sneddon:Book}. We accordingly write
\begin{equation}\label{fourier def}
\hat{\chi}(\bb{s},\bar{Z}) = \frac{1}{2\pi}\iint\chi(R,\bar{Z})e^{i\bb{s}\bcdot\bb{R}}\,\dd ^2\mathbf{R},
\end{equation}
where $\mathbf{R}$ is a position vector in the plane $\bar{Z}=0$, whose magnitude is $R$, and $\bb{s}$ is the corresponding transformation variable, whose magnitude is $s$. 

Similar to the analysis in the previous section, Laplace's equation \eqref{chi lap} and attenuation \eqref{chi decay} imply 
\begin{equation}\label{chi hat form}
\hat\chi(\bb{s},\bar{Z})= \frac{1}{s}Y(s)e^{-s\bar{Z}}.
\end{equation}
Taking the Fourier transform of the boundary condition \eqref{chi bc} using \eqref{chi hat form} yields
\begin{equation}\label{fourier transform bc}
\nabla^2_{\bb{s}}Y+2\left(\frac{1}{\alpha s}-1\right)Y=-\frac{4\pi}{\alpha}\delta_{\text{2D}}(\bb{s})
\end{equation}
where $\delta_{\text{2D}}(\bb{s})$ is the two-dimensional Dirac delta function.
Integrating \eqref{fourier transform bc} over a small circle of radius $s'$ and using the divergence law in the plane gives 
\begin{equation}\label{sing balance}
2\pi s' \left(\frac{\dd Y}{\dd s}\right)_{s=s'} +2\underset{s<s'}{\iint}\left(\frac{1}{\alpha s}-1\right)Y(s)\,d^2\bb{s} = -\frac{4\pi}{\alpha}
\end{equation}
In the limit $s'\to0$, the first and third term in \eqref{sing balance} balance to give
\begin{equation}\label{Y singular}
Y(s)\sim -\frac{2}{\alpha}\log s \quad \text{as} \quad s\to0,
\end{equation}
which \textit{a posteriori} justifies the neglect of the second term in \eqref{sing balance}. 

We can now restrict the problem governing $Y(s)$ to $s>0$, where \eqref{fourier transform bc} reduces to
\begin{equation}\label{Y reduced}
\frac{1}{s}\frac{\dd }{\dd s}\left(s\frac{\dd Y}{\dd s}\right)+2\left(\frac{1}{\alpha s}-1\right)Y=0,
\end{equation}
to be considered together with the singular boundary condition \eqref{Y singular} and the condition that $Y(s)$ attenuates fast enough for the transform to exist. 
Following the transformation $Y(s)=e^{-\sqrt{2}s}T(p)$, where $p=2\sqrt{2} s$, the governing equation \eqref{Y reduced} becomes
\begin{equation}\label{axi T eq}
p\frac{\dd ^2T}{\dd p^2}+(1-p)\frac{\dd T}{\dd p}+\tilde{n}T=0, \quad p>0,
\end{equation}
where the parameter $\tilde{n}$ is related to the scaled eigenvalue $\alpha$ through 
\begin{equation}\label{alpha n tilde}
\alpha = \frac{\sqrt{2}}{2\tilde{n}+1},
\end{equation}
while the asymptotic constraint \eqref{Y singular} becomes
\begin{equation}\label{T singular}
T(p)\sim -\frac{2}{\alpha}\log p \quad \text{as} \quad p\to0. 
\end{equation}

Misleadingly, \eqref{alpha n tilde} has the exact same form as the leading-order result \eqref{odd ev m pos} for $m=0$, with $\tilde{n}$ instead of $n$. Here, however, $\tilde{n}$ does not attain integer values. In fact, for any positive \emph{non-integer} $\tilde{n}$, a solution that satisfies \eqref{axi T eq}, \eqref{T singular} and has permissible behaviour at large $p$ is
\begin{equation}\label{axi T using U}
T(p)=\frac{2}{\alpha}\Gamma(-\tilde{n})U(-\tilde{n},1,p),
\end{equation}
where $U$ is the confluent hypergeometric function of the second kind; note that 
\begin{equation}\label{U asym}
U(-\tilde{n},1,p)\sim -\frac{1}{\Gamma(-\tilde{n})}\left[\log p +\Psi(-\tilde{n})+2\gamma\right]+o(1) \quad \text{as} \quad p\to0,
\end{equation}
$\Gamma(x)$ is the Gamma function and $\Psi(x)=\Gamma'(x)/\Gamma(x)$ is the Digamma function. 

At this stage $\tilde{n}$ remains nearly arbitrary, whereas we expect to find a discrete spectrum. Consider however the integral constrain \eqref{chi bc}, rewritten using \eqref{chi bc} as
\begin{equation}\label{int constraint solve}
\ln\frac{2}{h}+2\gamma=\frac{1}{2\pi}\iint\left(\alpha\pd{\chi}{\bar{Z}}+\frac{1}{H_0}\right)_{\bar{Z}=0}\,\dd^2{\bb{R}}
\end{equation}
The integral on the right hand side is most easily evaluated using the convolution identity for two-dimensional Fourier transforms \cite{Sneddon:Book}. Using \eqref{chi hat form} and noting that the transform of $1/H_0(R)$ is $2K_0(s\sqrt{2})$, where $K_0$ is the modified Bessel function of the second kind, we thereby find
\begin{equation}\label{convolution step}
\frac{1}{2\pi}\iint\left(\alpha\pd{\chi}{\bar{Z}}+\frac{1}{H_0}\right)\,\dd^2{\bb{R}}=\iint\left(-\alpha Y(s)+2K_0(s\sqrt{2})\right)\delta_{\text{2D}}(\bb{s})\dd^2\bb{s}.
\end{equation}
Using \eqref{axi T using U}, \eqref{U asym}, and noting that $K_0(x)\sim - \ln(x/2)-\gamma + o(1)$ as $x\to0$, \eqref{convolution step} yields
\begin{equation}
\underset{s\to0}{\lim} \left(-\alpha Y(s)+2K_0(s\sqrt{2})\right)=2\ln 4 + 2\Psi(-\tilde{n})+2\gamma.
\end{equation}

\begin{figure}[t!]
\begin{center}
\includegraphics[scale=0.35]{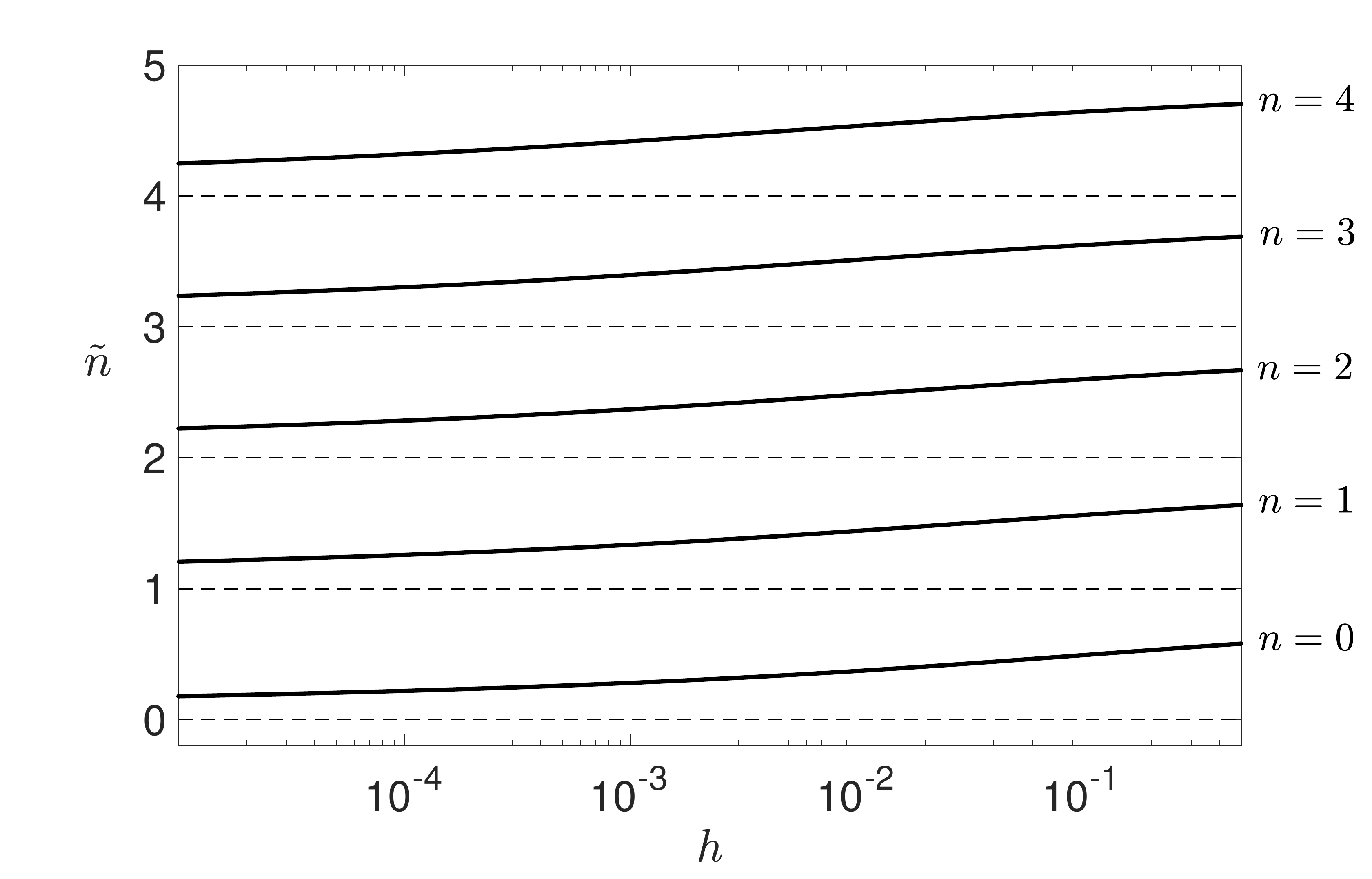}
\caption{First five solution branches of \eqref{trans} as a function of $h$. The equation is solved graphically by plotting $h=(1/8)\exp(-2\Psi(-\tilde{n}))$ as a function of $\tilde{n}$.}
\label{fig:trans}
\end{center}
\end{figure}
Thus we find from \eqref{int constraint solve} a transcendental equation for $\tilde{n}$:
\begin{equation}\label{trans}
2\Psi(-\tilde{n})=\ln\frac{1}{8h}.
\end{equation}
Clearly $\tilde{n}$ depends logarithmically upon $h$, with $\tilde{n}\to n$ as $h\to0$, where $n$ is a non-negative integer. The first few roots of \eqref{trans} are shown in figure \ref{fig:trans} as a function of $h$. From \eqref{alpha n tilde}, the scaled eigenvalues are
\begin{equation}\label{alpha n tilde 2}
\alpha = \frac{\sqrt{2}}{2\tilde{n}(n,\ln h)+1}, \quad n=0,1,2,\ldots,
\end{equation}
where for each $n$, $\tilde{n}$ is defined as the solution of \eqref{trans} that tends to $n$ as $h\to0$. We note that using the asymptotic behaviour of the Digamma function close to its poles it is easy to retrieve the first two terms in the logarithmic expansion \eqref{log series}, as well as higher-order ones.

\begin{figure}[t!]
\begin{center}
\includegraphics[scale=0.45]{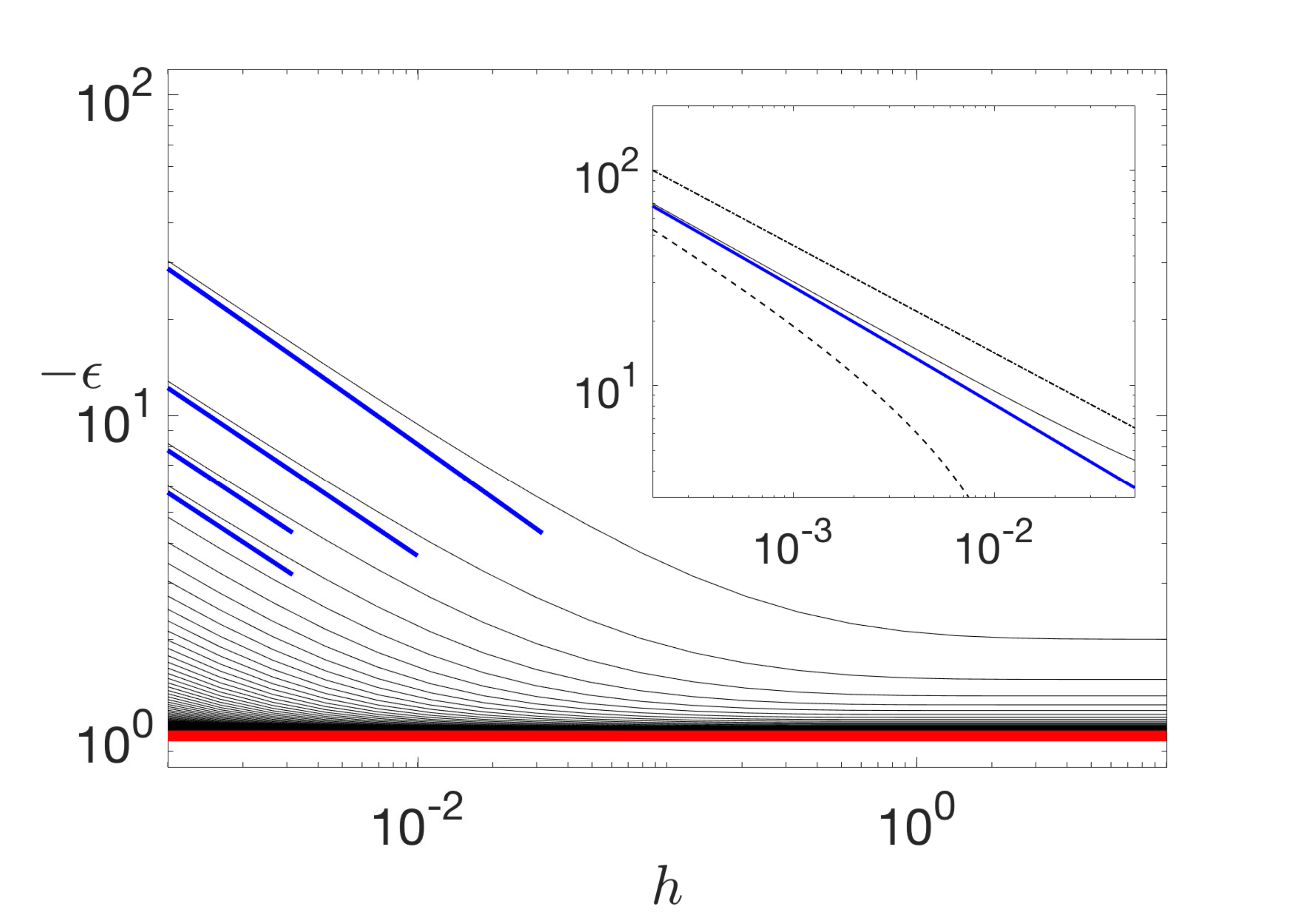}
\caption{Eigenvalues corresponding to axisymmetric ($m=0$) odd modes as a function of $h$, half the dimensionless gap width. The thick blue lines are the asymptotic predictions \eqref{axi epsi expansion}, with $\alpha$ given by \eqref{alpha n tilde 2} and $\tilde{n}(n,\ln h)$ determined from \eqref{trans} for $n=0,1,2,3$. The thin black lines are exact values obtained from the semi-numerical scheme described in \S\S\ref{ssec:exact}. The thick red line marks the accumulation point $\epsilon=-1$. The inset focuses on the fundamental mode $n=0$, showing in addition  one (dash-dotted line) and two (dashed line) terms of the logarithmic expansion \eqref{log series}.}
\label{fig:odd_m0}
\end{center}
\end{figure}
\subsection{Comparison with exact semi-analytical solutions}\label{ssec:comparisonodd_axi}
Figure \ref{fig:odd_m0} compares the asymptotic prediction \eqref{alpha n tilde 2} for the axisymmetric odd modes against the corresponding eigenvalues computed using the semi-analytical scheme discussed in \S\S\ref{ssec:exact}. The agreement is excellent for small $h$ and on par with the agreement found for $m\ne0$ in \S\ref{sec:odd}.

\subsection{Eigenfunctions in physical space}
\begin{figure}[t!]
\begin{center}
\includegraphics[scale=0.4]{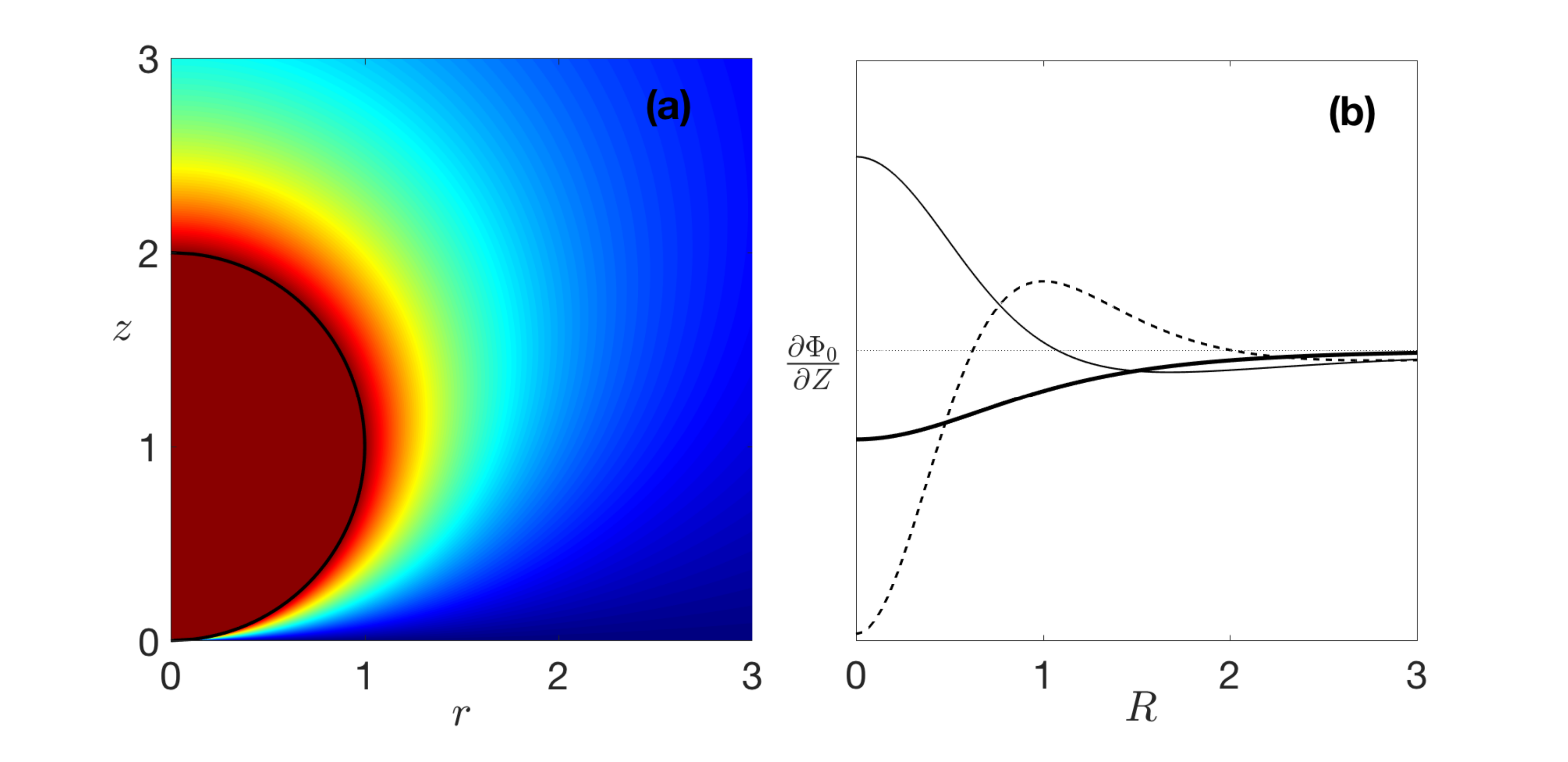}
\caption{Axisymmetric odd eigenfunctions from the asymptotic analysis of \S\ref{sec:axi}. (a) The outer potential  corresponds to nearly touching perfectly conducting spheres at opposite uniform potentials [cf.~\eqref{outer inside odd axi} and \eqref{outer sol}]. The outer potential is independent of the mode number $n$ and the small parameter $h$. (b) The transverse-field distribution in the gap region varies radially and depends on $n$ and $\ln h$ (here $h=0.001$); thick, thin and dashed lines depict modes $n=0,1$ and $n=2$, respectively [cf.~\eqref{axi gap field}].}
\label{fig:oddaxi_combined}
\end{center}
\end{figure}

The internal and external outer potentials \eqref{outer inside odd axi} and \eqref{outer sol}, which are independent of the mode number and $h$, are plotted in figure \ref{fig:oddaxi_combined}(a). In contrast, the gap and pole potentials depend on the mode number $n$ and $\ln h$ via the parameter $\tilde{n}$. Note that when inverting \eqref{fourier def} we may treat the two-dimensional Fourier transform as a Hankel transform:
\begin{equation}\label{chi transform hankel}
\chi(R,\bar{Z})=\underset{s\to R}{\mathcal{H}_0}\left(s^{-1}Y(s)e^{-s\bar{Z}}\right). 
\end{equation}
The field in the gap then follows from $A(R)H_0(R)=V\chi(R,0)+1$ as
\begin{equation}\label{axi gap field}
\frac{\dd \Phi_0}{\dd Z} = \frac{V}{H_0(R)}\left(1+\sqrt{2}(1+2\tilde{n})\Gamma(-\tilde{n})\int_0^\infty e^{-\sqrt{2} s} U(-\tilde{n},1,2\sqrt{2} s)J_0(Rs)\,ds\right).
\end{equation}
The first three field profiles are plotted in figure \ref{fig:oddaxi_combined}(b). 

\section{Even modes}\label{sec:even}
\subsection{Gap region} 
Consider now the eigenfunctions that are even about the plane $z=0$. In the gap region we pose the expansion 
\begin{equation}
\Phi \sim \Phi_0(R,Z)+h^{1/2}\Phi_{1/2}(R,Z)+h\Phi_1(R,Z) + \cdots.
\end{equation}
As we shall see, in the even case a leading-order solution entails consideration of the above three leading terms.
The $O(1)$, $O(h^{1/2})$ and $O(h)$ balances of Laplace's equation \eqref{gap laplace} respectively give
\begin{equation}\label{even lap orders}
\pd{^2\Phi_0}{Z^2}=0,\quad \pd{^2\Phi_{1/2}}{Z^2}=0, \quad \pd{^2\Phi_{1}}{Z^2}=-\frac{1}{R}\pd{}{R}\left(R\pd{\Phi_0}{R}\right)+\frac{m^2}{R^2}\Phi_0,
\end{equation}
whereas \eqref{sym gap even} yields at these orders the symmetry conditions
\begin{equation}\label{even sym orders}
\pd{\Phi_0}{Z}=0,\quad \pd{\Phi_{1/2}}{Z}=0, \quad \pd{\Phi_{1}}{Z}=0 \quad \text{at} \quad Z=0.
\end{equation}
It readily follows from \eqref{even lap orders} and \eqref{even sym orders} that $\Phi_0$ and $\Phi_{1/2}$ are independent of $Z$, i.e., $\Phi_0=\Phi_0(R)$ and $\Phi_{1/2}=\Phi_{1/2}(R)$; furthermore, integrating the equation for $\Phi_1$ with respect to $Z$ and using \eqref{even sym orders} yields
\begin{equation}\label{gap result}
\frac{1}{H_0}\left(\pd{\Phi_1}{Z}\right)_{Z=H_0(R)}=-\frac{1}{R}\frac{d}{dR}\left(R\frac{d\Phi_0}{dR}\right)+\frac{m^2}{R^2}\Phi_0.
\end{equation}

\subsection{Pole and outer regions and eigenvalue scaling}\label{ssec:evenscaling}
Continuity of potential \eqref{cont pole} suggests expanding the pole potential as 
\begin{equation}
\bar\Phi(R,\bar{Z})\sim \bar\Phi_0(R,\bar{Z}) + O(h^{1/2}).
\end{equation}
The scaling of the eigenvalue $\epsilon$ can now be extracted by considering the leading-order balance of the displacement-continuity condition \eqref{disp pole}:
\begin{equation}\label{even disp balance}
-\epsilon h^{-1/2}\pd{\bar\Phi_0}{\bar Z} \sim \frac{\dd H_0}{dR}\frac{\dd \Phi_0}{dR}-\pd{\Phi_1}{Z} \quad \text{at} \quad Z=H_0(R),\quad \bar{Z}=0.
\end{equation}
(Note that the independence of $\Phi_0$ upon $Z$ eliminates a potentially leading-order term proportional to $H_1(R)$.) 

The scaling 
\begin{equation}\label{even scaling}
\epsilon=O(h^{1/2})
\end{equation}
allows balancing the left-hand and right-hand sides of \eqref{even disp balance}. In this case, the projection of the radial field on the nearly transverse direction of the boundary normal, together with the comparably weak transverse field, balances the displacement associated with the strong transverse field in the pole region. Small $\epsilon$ implies that to leading order the \emph{external} outer potential satisfies a homogeneous Neumann boundary condition on the tangent-sphere boundary and hence must diverge as the gap is approached. The gap and pole potentials are therefore asymptotically large in comparison to the outer potential. 

Before proceeding with the analysis of the strongly localised even modes we note that \eqref{even scaling} is not the only possible scaling. Indeed, the regular scaling $\epsilon=O(1)$ leads to the anomalous even modes, which we will study in \S\ref{sec:anom}. Note that if $\epsilon$ is regular then \eqref{even disp balance} degenerates into a Neumann condition on $\bar\Phi_0$. 

\subsection{Effective eignevalue problem}
In light of the above discussion, the eigenvalue is expanded as
\begin{equation}\label{even epsi expansion}
\epsilon \sim -\alpha h^{1/2} + o(h^{1/2}),
\end{equation}
where $\alpha$ is a positive constant. As in the odd case, we shall develop an effective eigenvalue problem governing the prefactor $\alpha$ and the the pole potential $\bar\Phi_0$. From \eqref{lap pole}, the latter satisfies
\begin{equation}\label{lap pole even leading}
\pd{^2\bar\Phi_0}{\bar Z^2}+\frac{1}{R}\pd{}{R}\left(R\pd{\bar\Phi_0}{R}\right)-\frac{m^2}{R^2}\bar\Phi_0=0
\end{equation}
in the half space $\bar{Z}>0$, while \eqref{cont pole} and \eqref{disp pole} respectively yield the boundary conditions 
\begin{equation}\label{cont even}
\bar\Phi_0(R,0)=\Phi_0(R)
\end{equation}
and
\begin{equation}\label{disp even}
\alpha \pd{\bar\Phi_0}{\bar{Z}}=\frac{\dd H_0}{\dd R}\frac{d\Phi_0}{dR}-\pd{\Phi_1}{Z} \quad \text{at} \quad \bar Z=0,\quad {Z}=H_0(R).
\end{equation}
Using \eqref{gap result} and \eqref{lap pole even leading}, \eqref{cont even} and \eqref{disp even} are combined to give 
\begin{equation}\label{bar bc}
\alpha\pd{\bar\Phi_0}{\bar Z}+H_0\pd{^2\bar\Phi_0}{\bar{Z}^2}=R\pd{\bar\Phi_0}{R} \quad \text{at} \quad \bar Z=0,
\end{equation}
a boundary condition involving only $\bar\Phi_0$. Lastly, since the outer potential is asymptotically smaller than the pole potential, the latter satisfies the attenuation condition
\begin{equation}\label{even attenuate}
\bar\Phi_0\to0\quad \text{as} \quad R^2+\bar{Z}^2\to\infty.
\end{equation}

\subsection{Solution of the effective eigenvalue problem}
Consider the eigenvalue problem consisting of \eqref{lap pole even leading}, \eqref{bar bc} and \eqref{even attenuate}. Following the analysis in the odd case, we look for solutions in the form 
\begin{equation}\label{hankel odd even}
\bar{\Phi}_0(R,\bar{Z})=\underset{s\to R}{\mathcal{H}_m}\hat{\bar\Phi}(s,\bar{Z}). 
\end{equation}
Given \eqref{hankel odd even} and \eqref{even attenuate} we again write
\begin{equation}\label{Y def another}
\hat{\bar{\Phi}}=\frac{1}{s}Y(s)e^{-s\bar{Z}}
\end{equation}
with $Y(s)$ to be determined. To this end, transforming condition \eqref{bar bc} gives
\begin{equation}\label{bc t1}
\alpha Y(s) + \underset{R\to s}{\mathcal{H}_m}\left(R\pd{\bar{\Phi}_0}{R}\right)_{\bar{Z}=0}=sY(s)+\frac{1}{2}\underset{R\to s}{\mathcal{H}_m}\left(R^2\pd{^2\bar\Phi_0}{\bar{Z}^2}\right)_{\bar{Z}=0}.
\end{equation}
Assuming that $\bar\Phi_0(R,0)=o(1/R^2)$ as $R\to\infty$ we find by integrating by parts  
\begin{gather}\label{relation a}
\underset{R\to s}{\mathcal{H}_m}\left(R\pd{\bar{\Phi}}{R}\right)_{\bar{Z}=0}=-\frac{Y}{s}-\frac{dY}{ds}
\end{gather}
and
\begin{gather}\label{relation b}
\underset{R\to s}{\mathcal{H}_m}\left(R^2\pd{^2\bar\Phi}{\bar{Z}^2}\right)_{\bar{Z}=0} 
=-s\frac{\dd^2Y}{\dd s^2}-3\frac{\dd Y}{\dd s}-\frac{Y}{s}+\frac{m^2}{s}Y.
\end{gather}

By substituting \eqref{relation a} and \eqref{relation b} into \eqref{bc t1} we find 
\begin{equation}\label{Y eq}
s\frac{\dd^2Y}{\dd s^2}+\frac{\dd Y}{\dd s}-\left(\frac{1+m^2}{s}+2s-2\alpha\right)Y=0.
\end{equation}
Making the substitution 
\begin{equation}
Y(s)=s^{\sqrt{1+m^2}} e^{-\sqrt{2} s}T(p), 
\end{equation}
where $p=2\sqrt{2} s$, transforms \eqref{Y eq} into the associated Laguerre equation \eqref{assoc Lag odd} with parameters 
\begin{equation}
 \nu=2\sqrt{1+m^2}, \quad n = \frac{\alpha}{\sqrt{2}}-\frac{1}{2}-\sqrt{1+m^2}.
\end{equation}
Similar to the odd case, we conclude that $n$ must be a non-negative integer for the transform \eqref{hankel odd even} to exist and for $\Phi_0(R,0)$ to have the assumed attenuation rate as $R\to\infty$. The scaled eigenvalues are thus obtained as
\begin{equation}\label{alpha even}
\alpha = \sqrt{2}\left(n+\frac{1}{2}+\sqrt{1+m^2}\right), \quad n=0,1,2,\ldots,
\end{equation}
with associated eigenfunctions 
\begin{equation}\label{even Y}
Y(s)=s^{\sqrt{1+m^2}} e^{-\sqrt{2} s}L\ub{2\sqrt{1+m^2}}_n(2\sqrt{2} s), 
\end{equation} 
where we have chosen the multiplicative factor to be unity. 

\subsection{Eigenfunctions in physical space}
The eigenfunctions are readily inverted to give, for example, the radial distribution of the gap potential:
\begin{equation}\label{even gap potential}
\Phi_0(R) = \int_0^{\infty}s^{\sqrt{1+m^2}}e^{-s\sqrt{2}}L_n\ub{2\sqrt{1+m^2}}(2\sqrt{2}s)J_m(Rs)\,ds.
\end{equation}
It can be shown from \eqref{even gap potential} that $\Phi_0=O(1/R^{1+\sqrt{1+m^2}})$ as $R\to\infty$, except for $m=0$ in which case $\Phi_0=O(1/R^3)$. These attenuation rates are compatible with our assumption $\Phi_0 = o(1/R^{2})$ and imply that the outer potential is $O(h^{3/2})$ for $m=0$ and $O(h^{(1+\sqrt{1+m^2})/2})$ for $m\ne0$. 

\begin{figure}[t!]
\begin{center}
\includegraphics[scale=0.36]{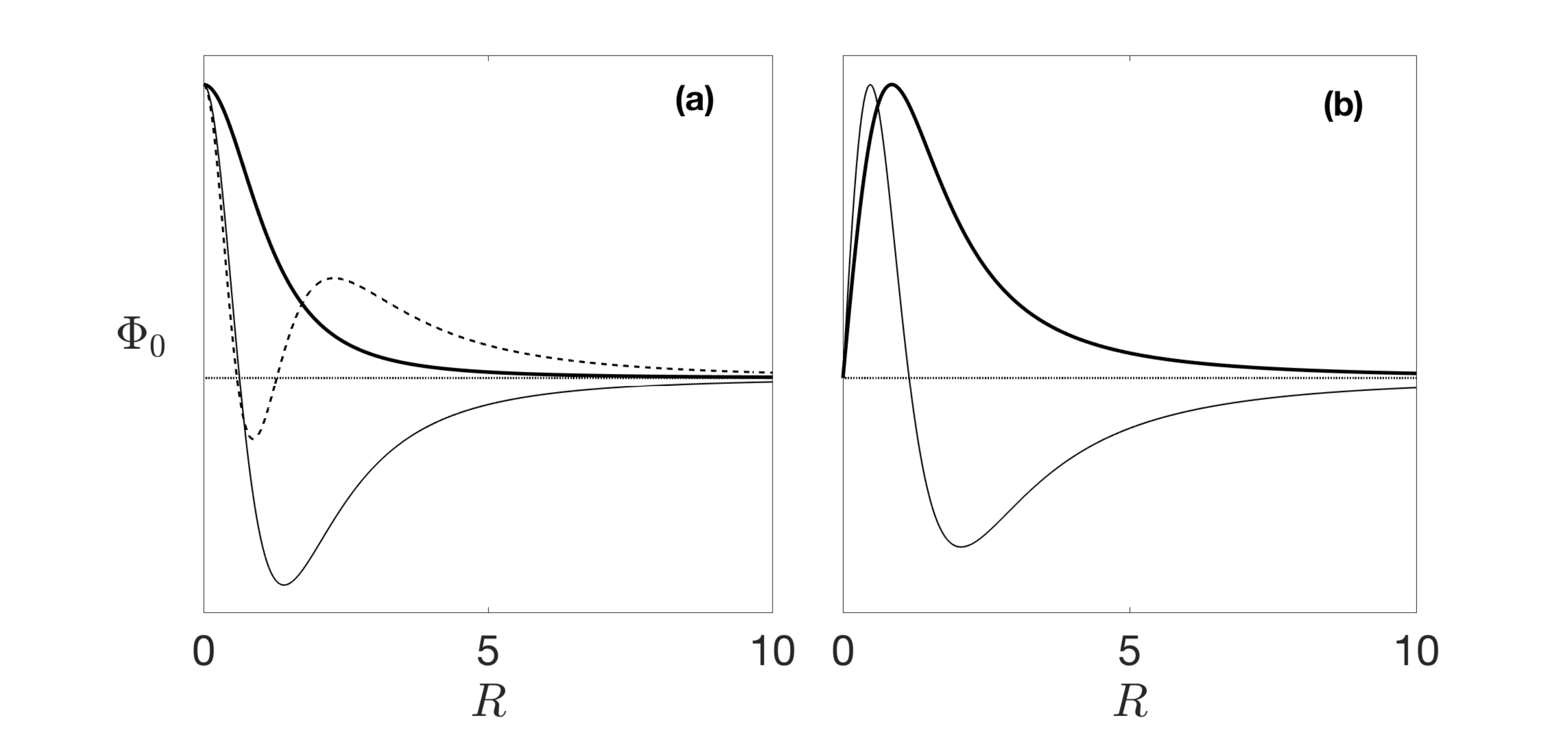}
\caption{Radial distributions \eqref{even pot m0} and \eqref{even pot m1} of the gap potentials for even modes with $m=0$ (a) and $m=1$ (b). Thick, thin and dashed lines depict modes $n=0,1$ and $n=2$, respectively.}
\label{fig:evengap_fields}
\end{center}
\end{figure}

The quadrature \eqref{even gap potential} can be evaluated exactly for given $m$ and $n$. For example, for $(m,n)=(0,\{0,1,2\})$: 
\begin{equation}\label{even pot m0}
\Phi_0(R) = \left\{\frac{\sqrt{2}}{(2+R^2)^{3/2}},\frac{\sqrt{2}(-2+5R^2)}{(2+R^2)^{5/2}},\frac{2\sqrt{2}(4+7R^2(R^2-2))}{(2+R^2)^{7/2}}\right\}.
\end{equation}
Similarly, for $(m,n)=(1,\{0,1\})$:
\begin{multline}\label{even pot m1}
\Phi_0(R) = \left\{2^{-2-\frac{1}{\sqrt{2}}}\Gamma(\sqrt{2}+2)R\, {_2F_1}\left(1+\frac{1}{\sqrt{2}},\frac{3}{2}+\frac{1}{\sqrt{2}},2,-\frac{R^2}{2}\right),\right. \\ \left.
2^{-\frac{5}{2}-\frac{1}{\sqrt{2}}}\Gamma(\sqrt{2}+2)R
\left[(4+\sqrt{2}){_2F_1}\left(1+\frac{1}{\sqrt{2}},\frac{3}{2}+\frac{1}{\sqrt{2}},2,-\frac{R^2}{2}\right) \right. \right. \\ \left. \left. - 4(1+\sqrt{2}){_2F_1}\left(\frac{3}{2}+\frac{1}{\sqrt{2}},2+\frac{1}{\sqrt{2}},2,-\frac{R^2}{2}\right)\right]
\right\}.
\end{multline}
We plot the gap potentials \eqref{even pot m0} and \eqref{even pot m1} in figure \ref{fig:evengap_fields}. 

\subsection{Comparison with exact semi-analytical solutions}\label{ssec:comparisoneven}
\begin{figure}[t]
\begin{center}
\includegraphics[scale=0.4]{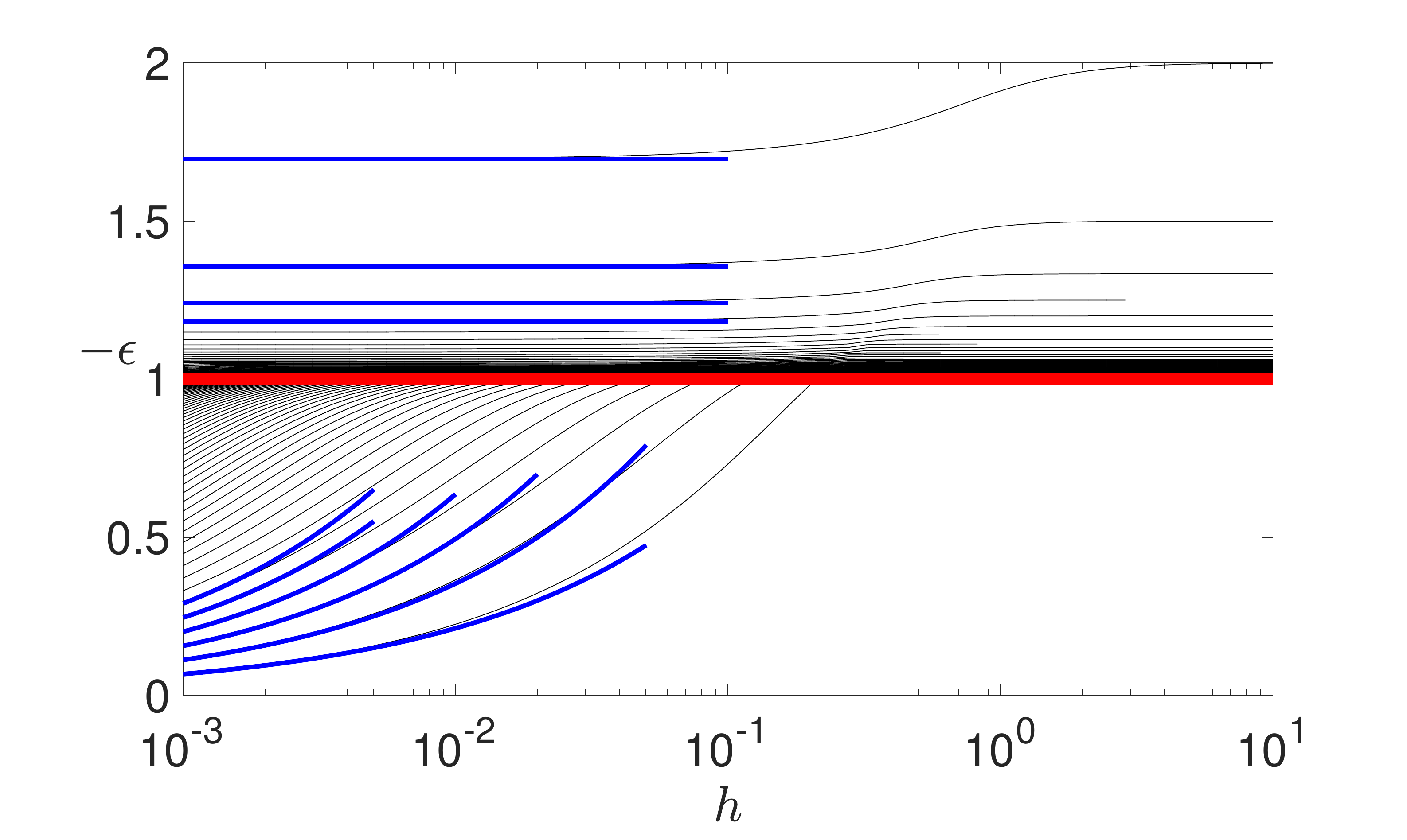}
\caption{
Eigenvalues corresponding to axisymmetric ($m=0$) even modes as a function of $h$, half the dimensionless gap width. The thick blue lines are the asymptotic predictions: localised gap modes ($-\epsilon<1$) --- \eqref{even epsi expansion} and \eqref{alpha even} for $n=0,1,\ldots,5$; anomalous modes ($-\epsilon>1$) --- \eqref{anom exp} and \eqref{anom asym}. The thin black  lines are exact values obtained from the semi-numerical scheme described in \S\S\ref{ssec:exact}. The thick red line marks the accumulation point $\epsilon=-1$.}
\label{fig:even_m0}
\end{center}
\end{figure}

\begin{figure}[h]
\begin{center}
\includegraphics[scale=0.4]{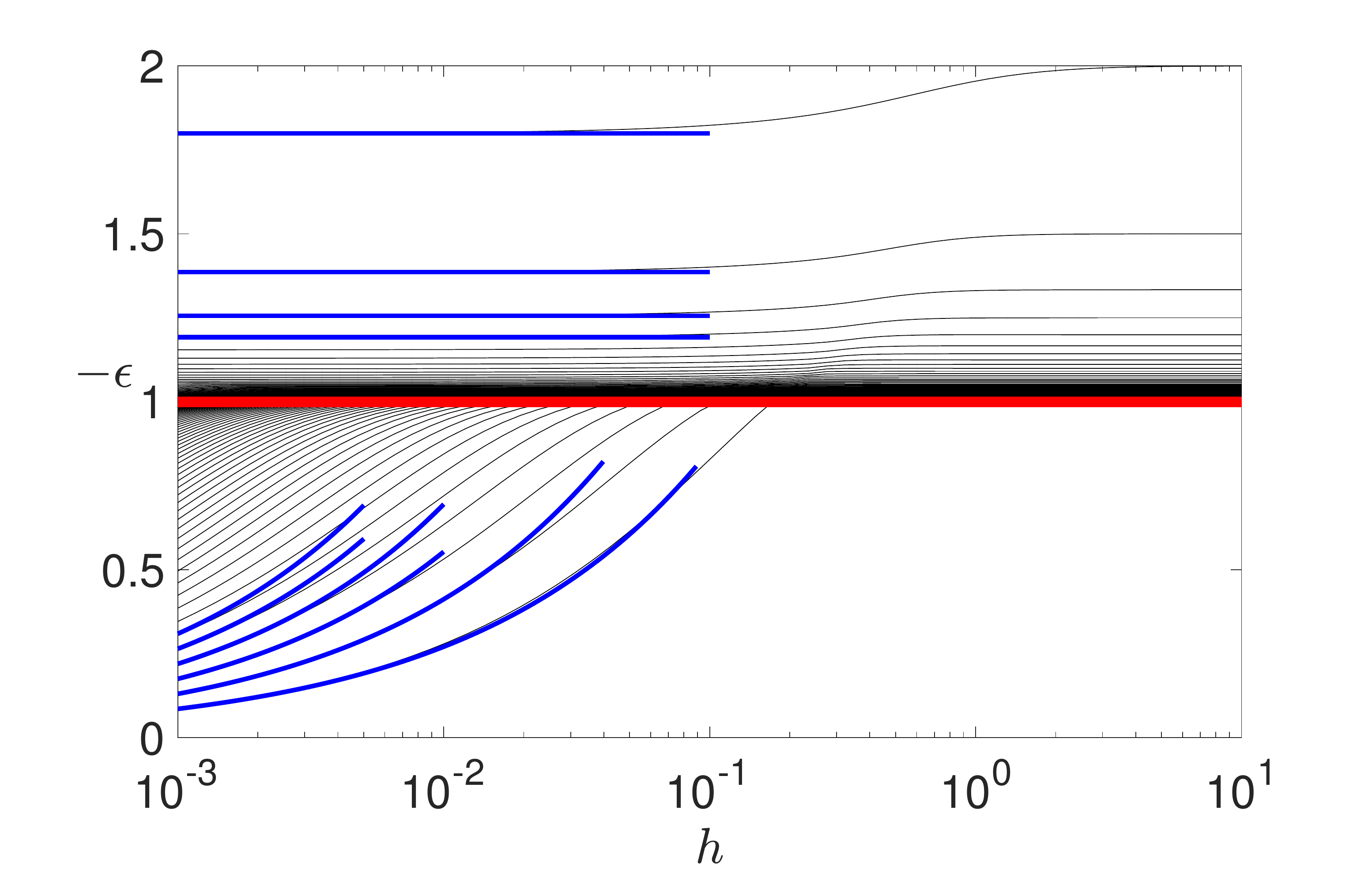}
\caption{Same as figure \ref{fig:even_m0} but for non-axisymmetric even modes with $m=1$.}
\label{fig:even_m1}
\end{center}
\end{figure}

\begin{figure}[h]
\begin{center}
\includegraphics[scale=0.4]{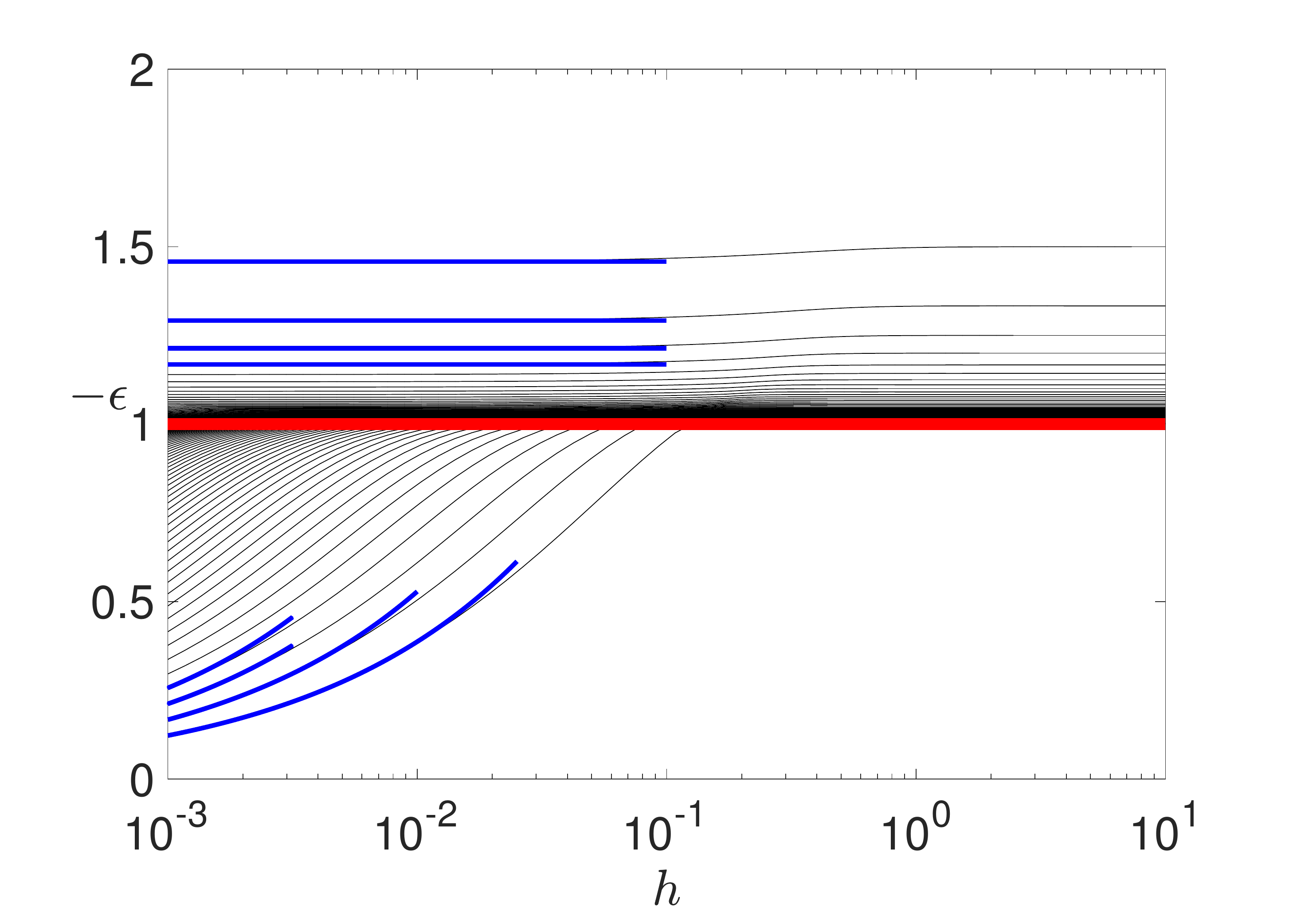}
\caption{Same as figure \ref{fig:even_m0} but for non-axisymmetric even modes with $m=2$.}
\label{fig:even_m2}
\end{center}
\end{figure}

Figures \ref{fig:even_m0}--\ref{fig:even_m2} present, respectively for $m=0,1$ and $2$, a comparison between the  asymptotic prediction \eqref{even epsi expansion}, with $\alpha$ given by $\eqref{alpha even}$, and the eigenvalues computed using the semi-analytical scheme discussed in \S\S\ref{ssec:exact}. 


\section{Anomalous even modes}\label{sec:anom}
\subsection{Scaling and outer eigenvalue problem}
We finally consider the anomalous even modes, for which 
\begin{equation}\label{anom exp}
\epsilon\sim \epsilon_0 +o(1) \quad \text{as} \quad h\to0,
\end{equation}
with $0>\epsilon_0=O(1)$. In this case, the leading-order eigenvalues and eigenfunctions will be determined by an effective problem confined to the outer region. 

Without loss of generality we now assume that the potentials are $O(1)$ in the outer region. We accordingly pose the expansions
\begin{equation}\label{anom expansion}
\bar{\psi}\sim {\bar\psi}_0(\xi,\eta)+o(1), \quad \psi\sim \psi_0(\xi,\eta)+o(1),
\end{equation} 
where $(\xi,\eta)$ are the tangent-sphere coordinates introduced in \S\S\ref{ssec:outer}. The leading-order potentials satisfy Laplace's equation
\begin{equation}\label{anom lap}
\nabla^2\bar\psi_0=0 \quad \text{for} \quad 0<\xi<1; \quad \nabla^2\psi_0=0  \quad \text{for} \quad \xi>1;
\end{equation}
the continuity condition
\begin{equation}\label{anom cont}
{\bar\psi}_0={\psi}_0 \quad \text{at} \quad \xi=1;
\end{equation}
and displacement condition, 
\begin{equation}\label{anom disp}
\epsilon_0\pd{{\bar\psi}_0}{\xi}=\pd{{\psi}_0}{\xi} \quad \text{at} \quad \xi=1;
\end{equation}
the symmetry condition 
\begin{equation}\label{anom sym}
\pd{\psi_0}{\xi}=0 \quad \text{at} \quad \xi=0;
\end{equation}
and attenuation
\begin{equation}\label{anom decay}
\psi_0=0 \quad \text{as} \quad \xi^2+\eta^2\to0. 
\end{equation}

The above eigenvalue problem is closed by matching considerations in the limit $\eta\to\infty$. For $\epsilon=O(1)$ the pole potential approximately satisfies a homogeneous Neumann condition at $\bar{Z}=0$ (see \S\S\ref{ssec:evenscaling}) and thus forced solely by matching with the internal outer potential. For $m=0$, these conditions are consistent with a uniform leading-order pole (and gap) potential. The requisite matching condition for $m=0$ is thus 
\begin{equation}\label{regularity axi}
\psi_0\to \text{const.} \quad \text{as} \quad \eta\to\infty.
\end{equation}
A uniform pole potential is obviously impossible for $m\ne0$. Instead, the leading-order pole solution must grow as $R^2\to\bar{Z}^2\to\infty$; hence the outer solution is asymptotically large compared to the potentials in the gap and pole regions. The requisite matching condition for $m\ne0$ is thus 
\begin{equation}\label{regularity nonaxi}
\psi_0\to 0 \quad \text{as}\quad  \eta  \to\infty.
\end{equation}

\subsection{Solution of the outer eigenvalue problem}\label{ssec:anomsol}
A general solution that satisfies Laplace's equations \eqref{anom lap}, the continuity condition \eqref{anom cont}, the symmetry condition \eqref{anom sym}, and  attenuation \eqref{anom decay} can be obtained by superposing solutions of Laplace's equation in tangent-sphere co-ordinates \cite{Moon:61}:
\begin{gather}\label{anom sol in}
{\bar{\psi}}_0(\xi,\eta)=
 (\xi^2+\eta^2)^{1/2}\underset{s\to\eta}{\mathcal{H}_m}[B(s)s^{-1}e^{s}\cosh(s)e^{-\xi s}],\\ \label{anom sol out}
\psi_0(\xi,\eta) =(\xi^2+\eta^2)^{1/2}\underset{s\to\eta}{\mathcal{H}_m}
[B(s)s^{-1}\cosh(s\xi)],
\end{gather}
where $B(s)$ is to be determined subject to the remaining conditions and existence of the transforms. 

Transforming the displacement condition \eqref{anom disp} in conjunction with \eqref{anom sol in} and \eqref{anom sol out}, followed by application of identity \eqref{Identity}, yields: 
\begin{multline}\label{Beq}
s^2\left(\epsilon_0\cosh s+\sinh s\right)\frac{\dd^2B}{\dd s^2}+s\left(2s\epsilon_0\sinh s+(2s+\epsilon_0)\cosh s+\sinh s\right)\frac{\dd B}{\dd s}\\ +\left[\epsilon_0(s-m^2)\cosh s+(s\epsilon_0-m^2)\sinh s\right] B=0.
\end{multline}
In the limit $s\to0$ solutions of \eqref{Beq} with $m=0$ are easily shown to be $O(1)$ or $O(\log s)$, whereas for $m\ne0$ the solutions are $O(s^m)$ or $O(s^{-m})$. Only the regular solutions are compatible with our use of identity \eqref{Identity}, which suggests rejecting the singular ones. The same conclusion also follows from the respective matching conditions \eqref{regularity axi} and \eqref{regularity nonaxi}. Indeed, by inverting \eqref{anom sol out} it can be shown that 
\begin{equation}\label{B invert}
B(s)\sim \int_0^{\infty}\frac{\psi_0(\xi,\tau/s)}{(s^2\xi^2+\tau^2)^{1/2}}J_m(\tau)\tau\,\dd\tau \quad \text{as} \quad s\to0,
\end{equation}
whereby using \eqref{regularity axi} and \eqref{regularity nonaxi} we find that $B(s)=O(1)$ for $m=0$ and $B(s)=o(1)$ for $m\ne0$. 

\begin{figure}[t]
\begin{center}
\includegraphics[scale=0.38]{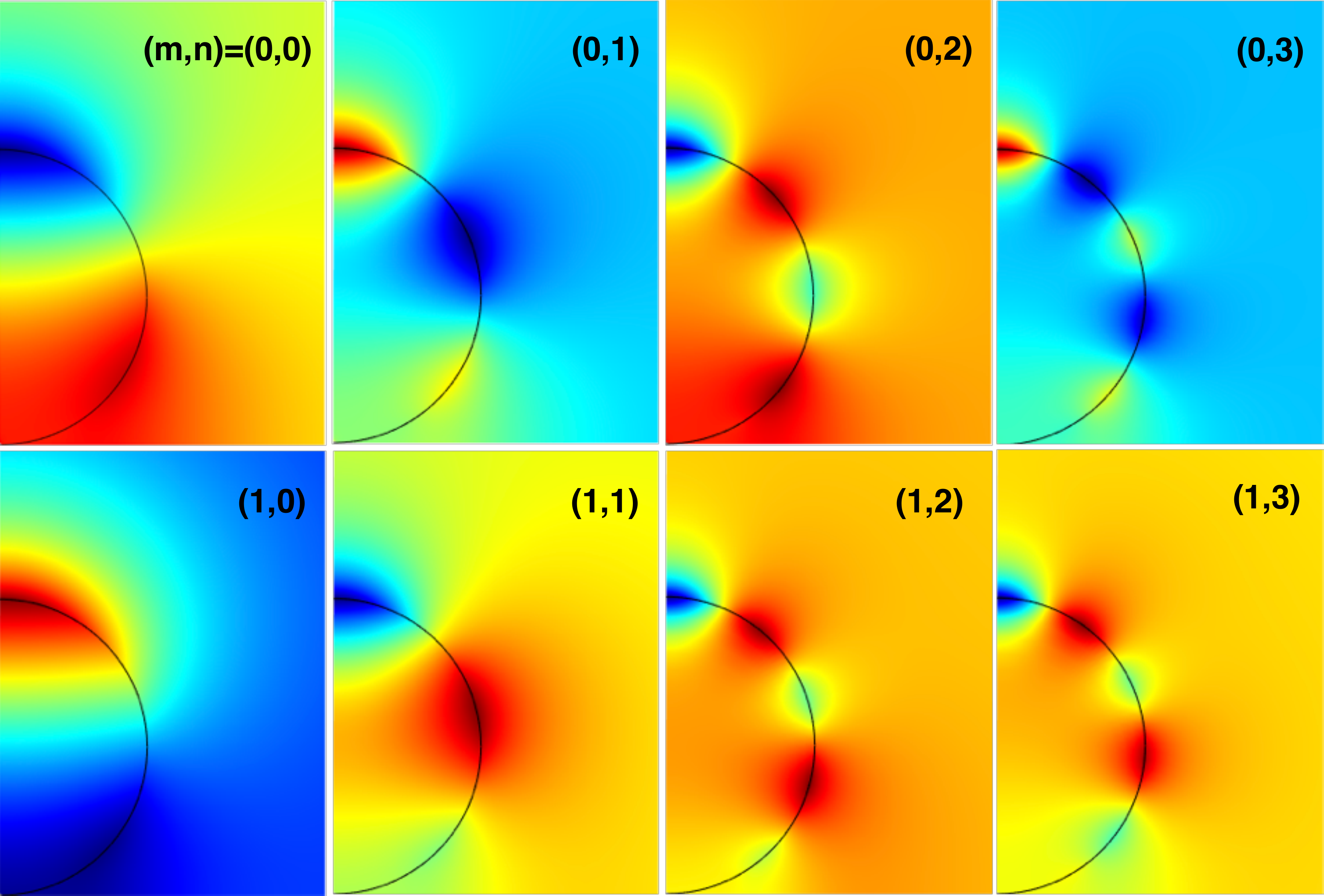}
\caption{Leading-order outer potentials of anomalous even modes with $(m,n)$ as shown. In the gap and pole regions the potentials are uniform for $m=0$ and asymptotically small for $m\ne0$. The colours red and blue respectively mark maximum and minimum values.}
\label{fig:anom}
\end{center}
\end{figure}
In light of the above, a higher-oder analysis of \eqref{Beq} in the limit $s\to0$ shows that 
\begin{equation}\label{B small s}
B\sim \text{const.} \times \left(s^m-\frac{2m+\epsilon_0}{(1+2m)\epsilon_0}s^{m+1}+\cdots\right) \quad \text{as} \quad s\to0
\end{equation}
for any positive integer $m$, where given the multiplicative freedom the constant prefactor can be chosen arbitrarily. For any given $\epsilon_0$, the small-$s$ behaviour \eqref{B small s} determines the solution to \eqref{Beq}. An asymptotic analysis of \eqref{Beq} in the latter limit \cite{Bender:13}, however, yields solutions attenuating like $s^{-1/(1+\epsilon)}\exp(-2s)$ and $s^{-\epsilon/(1+\epsilon)}$ and clearly only the former is acceptable. Thus $B(s)$ must satisfy both \eqref{B small s} and exponentially decay, which is only possible for special values of $\epsilon_0$. We solve this eigenvalue problem using a shooting method, where we integrate backwards starting from the exponentially decaying solution at large $s$, choosing $\epsilon_0$ such that \eqref{B small s} is satisfied as $s\to0$. Using this scheme we calculated the following eigenvalues:
\begin{align}\label{anom asym}
m=0: &\quad \epsilon \sim \left\{-1.6964,-1.3553,-1.2412,-1.1837,\ldots\right\};  \nonumber \\ 
m=1: &\quad \epsilon \sim \left\{-1.7999,-1.3862,-1.2562,-1.1926,\ldots \right\};  \nonumber \\ 
m=2: &\quad \epsilon \sim \left\{-1.4582, -1.2918,-1.2138,-1.1689,\ldots\right\},
\end{align}
and corresponding eigenfunctions $B(s)$, 
where as usual for each $m$ the eigenvalues are ordered with increasing closeness to the accumulation point. 

Excellent agreement between \eqref{anom asym} and the eigenvalues computed using the semi-analytical scheme discussed in \S\S\ref{ssec:exact} is presented in figure \ref{fig:even_m0}--\ref{fig:even_m2}. Figure \ref{fig:anom} depicts the first four modes for $m=0$ and $m=1$.

\section{Approximations in the literature}\label{sec:comparison}
As already noted, the sphere-dimer geometry has been extensively studied using the infinite algebraic system obtained from separation of variables in bi-spherical coordinates, or using transformation optics followed by separation of variables in spherical coordinates. In particular, this exact formulation has been used as a starting point for deriving approximations in the near-contact limit. It is useful to note in which cases these approximations coincide with the ones derived here using matched asymptotic expansions. 

Consider first the odd modes. Our leading-order approximation for the eigenvalues $\eqref{odd ev m pos}$ is equivalent to the approximation obtained by Klimov and coworkers \cite{Klimov:07,Klimov:07cluster,Klimov:14}. As we have seen, this approximation is only accurate for the non-axisymmetric odd modes ($m\ne0$). The fact that there is a logarithmic error in the axisymmetric case $m=0$ was first noted by Lebedev \textit{et al.} \cite{Lebedev:10,Lebedev:13}. Our algebraically accurate approximation for $m=0$ \eqref{alpha n tilde 2} includes all the terms in an infinite expansion in inverse logarithmic powers. Incidentally, \eqref{log series} reveals that the leading logarithmic correction given in \cite{Lebedev:13} is off by a factor of two. 
Consider next the gap-localised even modes. Our approximation for the eigenvalues \eqref{alpha even} is equivalent to the approximation obtained by Klimov and coworkers \cite{Klimov:07cluster,Klimov:14}, only that their result is given in terms of an infinite series. We now see that the latter series is nothing but the expansion of $\sqrt{1+m^2}-m$ about $m=\infty$. Finally, we note that we have not found analytical approximations in the literature for the anomalous even modes. Pendry \textit{et al.} provide a heuristic formula that in the axisymmetric case can be fitted to give good agreement with computed eigenvalues \cite{Pendry:13}. 

\section{Concluding remarks}\label{sec:conc}
We have obtained asymptotic approximations for all the plasmonic eigenvalues and eigenfunctions of a sphere dimer in the near-contact limit. Using the spectral decomposition method for localised-surface-plasmon resonance, these results could be used to study the resonant response of a pair of closely spaced nanometallic spheres to arbitrary external forcing and for arbitrary material parameters (with ohmic losses accounted for). While this is outside the scope of this paper, we note that calculating the response in this way entails normalising the eigenmodes and evaluating the overlap between the excited eigenmodes and the external radiation. The description of the eigenmodes in the form of matched asymptotic expansions is particularly suitable for this purpose. Indeed, in many cases the associated  integrals would be confined to the gap and pole regions, where the geometry and form of the solution are greatly simplified; in fact, the integration could be carried out directly in Hankel space by using the Plancherel theorem \cite{Sneddon:Book} (for an example, see \cite{Schnitzer:15plas}). Even without carrying out detailed calculations, the scalings and asymptotic structure we found could be used to rapidly estimate scattering cross sections, localisation and field enhancements, thus providing significant insight to plasmonic resonance with nearly touching particles. 
 
Our approach could be generalised to related closely spaced geometries, such as dissimilar spherical particles, a particle close to a plane substrate, and in contrast to other analytical approaches, also non-spherical particle pairs. For modes strongly localised to the gap and pole regions, it is clear that the geometry would only come in through the curvatures at the point of minimum separation. For modes that involve the outer region to leading order, the outer problem could be reduced via matching to a regularised problem that is amenable to a straightforward numerical solution. 
Further generalisations of interest would be clusters of more than two closely spaced plasmonic particles \cite{Klimov:07cluster,Chuntonov:11} and periodic arrangements of nearly touching particles \cite{Sukharev:07,Klimov:14}. Matched asymptotic expansions could also be used to study other types of nearly singular geometries such as elongated nanorods \cite{Chen:13}. 

It is important to emphasise that in this paper we analysed the near-contact limit with the mode numbers $m$ and $n$ fixed. It is evident that the near-contact and high-mode-number limits do not commute: in the former limit the eigenvalues tend to either negative infinity, zero or constants smaller than $-1$, whereas in the latter all of the eigenvalues approach the accumulation point $-1$. Intuitively, a high mode number implies fast oscillations along and exponential decay away from boundaries. When taking that limit with the sphere-dimer geometry fixed, the modes approach the high-mode-number modes of isolated spheres and accordingly $\epsilon\sim-1-1/n$. A more interesting distinguished limit is $n,m=O(1/h^{1/2})$, where the interaction between the spheres is important yet our present analysis breaks down. Approximations have been suggested in the limit of high azimuthal number $m$ based on the bi-spherical scheme \cite{Klimov:14}; it has been shown that these modes are important when calculating the van der Waals attraction between spheres \cite{Klimov:09,Pendry:13}. It would be interesting to consider limits where either or both $n$ and $m$ are large using singular perturbation techniques. This will necessarily involve not only matched asymptotic expansions but also WKBJ theory \cite{Hinch:91}, as the plasmon wavelength will be small compared with local radii of curvature.

We finally recall that our analysis is based on a quasi-static formulation valid in the limit where the structure is small compared to the wavelength. This is often the preferable scenario in nanoplasmonics, since for larger particles plasmon resonances are usually damped by radiation losses. Nevertheless, radiation corrections are often large enough to be of practical interest. Moreover, the singular near-contact and quasi-static limits probably do not commute, suggesting that in some cases retardation may play a role even for small structures. Remarkably, the plasmonic eigenvalue problem and its concomitant spectral theory can be generalised to the full Maxwell equations with a Silver-M\"uller radiation condition applied at large distances \cite{Bergman:80,Agranovich:99,Klimov:14,Ammari:16,Farhi:16,Chen:17}. In this formulation, which has several advantages \cite{Agranovich:99,Chen:17} over expansions in quasi-normal modes \cite{Sauvan:13}, the resonant permittivity values become frequency dependent and, owing to radiation losses, complex valued.

\textbf{Acknowledgements.} The author acknowledges funding from EPSRC New Investigator Award EP/R041458/1.

\bibliography{refs.bib}
\end{document}